\documentclass[aps,prd,twocolumn,groupedaddress,nofootinbib]{revtex4-1}

\usepackage{graphicx,color,amsmath}
\usepackage[export]{adjustbox}
\usepackage{slashed}
\usepackage{booktabs} 
\usepackage{braket}
\usepackage{hyperref}

\usepackage{tikz}
\usepackage{tkz-euclide}
\usetikzlibrary{decorations.pathmorphing}	
\tikzset{
    v/.style={decorate, decoration={snake, segment length=3mm, amplitude=0.75mm}, draw},
    f/.style={draw=black, postaction={decorate},
        decoration={markings,mark=at position .6 with {\arrow[very thick]{latex}}}},
    fb/.style={draw=black, postaction={decorate},
        decoration={markings,mark=at position .4 with {\arrowreversed[very thick]{latex}}}},
    fnar/.style={draw=black},
    g/.style={decorate, draw=black,
        decoration={coil,amplitude=3pt, segment length=3.5pt}},
    s/.style={dashed,draw=black, postaction={decorate},
        decoration={markings,mark=at position .55 with {\arrow[very thick]{latex}}}},
    sb/.style={dashed,draw=black, postaction={decorate},
        decoration={markings,mark=at position .55 with {\arrowreversed[draw=black,very thick]{latex}}}},
    snar/.style={dashed,draw=black,line width =1.25pt},
}

\newcommand{\MeV}{{\, {\rm MeV}}}
\newcommand{\GeV}{{\, {\rm GeV}}}
\newcommand{\TeV}{{\, {\rm TeV}}}

\newcommand{\fbm}{{\, {\rm fb^{-1}}}}

\newcommand{\Am}{{\mathcal{A}}}
\newcommand{\LL}{{\mathcal{L}}}
\newcommand{\MM}{{\mathcal{M}}}
\newcommand{\OO}{{\mathcal{O}}}

\definecolor{mypurple}{RGB}{164,64,214}

\newcounter{qnumber}

\begin{document}

\title{Dark forces coupled to non-conserved currents}

\author{Jeff A. Dror}
\email{ajd268@cornell.edu}
\affiliation{Department of Physics, LEPP, Cornell University, Ithaca, NY 14853}
\author{Robert Lasenby}
\email{rlasenby@perimeterinstitute.ca}
\affiliation{Perimeter Institute for Theoretical Physics, 31 Caroline Street N, Waterloo, Ontario N2L 2Y5, Canada}
\author{Maxim Pospelov}
\email{mpospelov@perimeterinstitute.ca}
\affiliation{Perimeter Institute for Theoretical Physics, 31 Caroline Street N, Waterloo, Ontario N2L 2Y5, Canada}
\affiliation{Department of Physics and Astronomy, University of Victoria, Victoria, BC V8P 5C2, Canada} 
\date{\today}

\begin{abstract}
	New light vectors with dimension-4 couplings to Standard Model states have 
	${\rm (energy / vector~mass)}^2$ enhanced production rates unless the current they couple to is conserved.
	These processes allow us to derive new constraints on the couplings of such
	vectors, that are significantly stronger than the previous literature
	for a wide variety of models.
	Examples include vectors with axial couplings to quarks
	and vectors coupled to currents (such as baryon number) that
	are only broken by the chiral anomaly.
	Our new limits arise from a range of processes, including
	rare $Z$ decays and flavor changing meson decays, and rule out 
	a number of phenomenologically-motivated proposals.
\end{abstract}

\maketitle

\section{Introduction}

New states beyond the Standard Model (SM) may have gone undetected
either because they are too heavy to be produced in large numbers at
collider experiments, such as those proposed by most solutions to the hierarchy problem, or because
they couple very weakly to the SM states. In the latter case,
such particles may be light enough to be produced
in experiments other than the highest-energy colliders, 
and can have a diverse range of experimental signatures~\cite{Hewett:2012ns,Essig:2013lka,Alexander:2016aln}.
In this paper, we will focus on new light vector particles;
these have been discussed extensively in the existing literature,
for purposes including
addressing experimental anomalies at low
energies~\cite{Gninenko:2001hx,Kahn:2007ru,Pospelov:2008zw,TuckerSmith:2010ra,Batell:2011qq,Feng:2016ysn}, explaining puzzles such as
baryon stability~\cite{Carone:1994aa}, or acting as a mediator to a dark
sector~\cite{Boehm:2003hm,Pospelov:2007mp,ArkaniHamed:2008qn}.

For a light vector with dimension-4 couplings to the Standard Model,
unless the SM current that the vector couples
to is conserved,
there are processes with ${\rm (energy / vector~mass)}^2$ rates involving
the longitudinal mode of the new vector.
In many such models,
these energy-enhanced processes can be the dominant production
mechanism in high-energy experiments, and can place
strong constraints on the vector's coupling.
A number of
works~\cite{Kahn:2007ru,Fayet:2006sp,Karshenboim:2014tka,Barger:2011mt}
have used enhanced longitudinal production to place constraints
on vectors with axial couplings to SM fermions. We extend these in a
variety of ways. 
For vectors with axial, generation-non-universal, or
$SU(2)_L$-violating couplings to SM fermions, we identify processes
which yield stronger constraints than those in previous works.
Most significantly, flavor-changing neutral current processes
involving the new vector can be enhanced
by $({\rm weak~scale}/{\rm vector~mass})^2$ compared to
competing processes. The resulting constraints 
are often the most powerful available, for vectors below the $B$ mass.

We also point out that if the new vector couples to a current
that is conserved at tree level, but broken by the chiral
anomaly (for example the SM baryon number current), this anomalous
non-conservation still gives rise to energy-enhanced longitudinal mode
emission. 
These loop-level, but ${\rm (energy / vector~mass)}^2$ enhanced, processes
can place significantly stronger constraints on light vectors than existing `tree-level' constraints. 
In the absense of fine-tuning, they can only be avoided at the expense of introducing
extra sources of electroweak symmetry breaking
in the UV theory, such as new sets of SM-chiral fermions.
Such options generally run into strong experimental constraints.
Conversely, cancelling the anomalies with new heavy fermions,
that obtain their masses from a SM-singlet vacuum expectation value (VEV), always results in
enhanced longitudinal emission.
These points are also discussed, more concisely, in an accompanying
letter~\cite{Dror:2017ehi}.

Turning to the structure of the paper, in section~\ref{sec:anom}, we discuss current non-conservation through
the chiral anomaly, while in section~\ref{sec:tree}, we treat
more general non-conserved currents.
Section~\ref{sec:phenomodels} describes how our new bounds
can constrain several models of phenomenological interest. In particular, we show that several proposals to explain the $^8$Be anomaly~\cite{Krasznahorkay:2015iga} are effectively ruled out. Section~\ref{sec:directions}
discusses some of the new experimental searches motivated by our analyses.


\section{Anomalous vectors}
\label{sec:anom}

Even if the current that a vector couples to is conserved
at tree level, it may be broken by chiral anomalies (examples in the
SM are the baryon and lepton number currents). 
As a result, a vector $X$ coupled to such a current
results in a non-renormalisable SM + $X$ effective field theory (EFT)
--- either the $U(1) _X$ symmetry is broken, or there is additional electroweak
symmetry breaking (EWSB). In the former case, the UV cutoff scale
of the theory is $\sim m_X/g_X$ (or below).
This non-renormalisability is reflected in the divergence at high energies
of some amplitudes in the EFT, involving fermion
triangle diagrams (and box diagrams) with longitudinal
$X$ modes. In this section, we review how to compute these amplitudes, and show how the anomaly results in enhanced longitudinal mode ($X_L$) emission
in various circumstances.

This picture can appear puzzling from the perspective of a UV theory,
in which the anomalies can be cancelled by new heavy fermions.
The resolution is that, in addition to the fermion-mass-independent
`anomalous'~\cite{AdlerAnomalies} part of loop amplitudes, there
is a piece that depends on the mass of the fermions in the loop.
In the limit where the new
anomaly-cancelling fermions have large masses compared to the external
momenta around the loop, their contribution will correspond to that
expected in the anomalous low-energy EFT, as we show explicitly below. Conversely, in the limit where their masses are small compared to the typical energies involved in the process, the mass-dependent piece vanishes, resulting in the usual anomaly-free result. 

These considerations have been discussed in a number of previous
papers, including~\cite{Anastasopoulos:2006cz,Dedes:2012me,Arcadi:2017jqd,Ismail:2017ulg},
as well as the original papers of D'Hoker and Farhi~\cite{DHoker:1984izu,DHoker:1984mif}. Our contribution in this section is essentially to present clearly the relation between
the UV physics and the low-energy theory, and to point out that
a $U(1)_X$-breaking theory results in energy-enhanced longitudinal production
of a light $X$ via anomalous couplings.


\subsection{Vectorial couplings}

Since fermion masses introduce non-conservation of axial currents, a
vector coupled to a tree-level-conserved current must have vectorial
couplings to the SM fermions (and since the SM Yukawa couplings are
non-diagonal, these couplings must be generation-universal).
As the photon and gluon also have vectorial couplings,
there is no mixed anomaly between the new vector $X$ and
QED or QCD. However, the chiral anomaly will generically 
lead to non-conservation of the $U(1)_X$ current,
\begin{equation}
	\partial^\mu J_\mu^X = \frac{\Am_{XBB}}{16 \pi^2} 
	\left( g'^2
	B_{\mu\nu} \tilde{B}^{\mu\nu} - g^2 W^a_{\mu\nu} (\tilde{W}^a)^{\mu\nu} \right)\,,
\end{equation}
where $\mathcal{A}_{XBB}
\equiv \text{Tr} \left[ Q _X Y ^2 \right]$,
and 
$\Am _{ XWW} \equiv \text{Tr} \left[ Q _X T ^a T ^a \right] = -\mathcal{A}_{XBB}$ 
(since the current is vectorial), with the traces taken over
the SM fermions and where $Q_X$ are their $U(1)_X$ charges.
$V ^{ \mu\nu } $ are the
field strength tensors, and $ \tilde{V} ^{ \mu \nu } \equiv \frac{1}{2}
\epsilon ^{ \mu \nu \sigma \rho } V _{ \sigma \rho } $.
If $\Am_{XBB}$ is non-zero, then the SM + $X$ EFT is not 
renormalisable, and some amplitudes will diverge at high energies.

In general, the effective theory 
may break the electroweak and U(1)$ _ X $ symmetries and hence we can include 
dimension-4 Wess-Zumino (WZ) terms,
\begin{align}
\label{eq:WZ}
	\LL & \supset C_B g_X g'^2 \epsilon^{\mu\nu\rho\sigma} X_\mu B_\nu \partial_\rho B_\sigma  \notag \\ 
	& + C_W g_X g^2 \epsilon^{\mu\nu\rho\sigma} X_\mu (W^a_\nu \partial_\rho W^a_\sigma + \frac{1}{3} g \epsilon^{abc} W^a_\nu W^b_\rho W^c_\sigma)\,.
\end{align}
There are multiple ways of evaluating anomalous amplitudes
within the effective theory, corresponding to different regularization
schemes. The combination of a given regularization scheme, and
particular values for the WZ coefficients, fixes the behaviour
of the theory.
To avoid breaking the electromagnetic gauge
symmetry, we must have $C_B = - C_W$.\footnote{EM gauge invariance also forbids a $\epsilon B W \partial W$ term. However,
such a term will appear in the low-energy theory obtained by integrating out
e.g.\ the top quark, in which the fermion content gives a chiral anomaly
between the SM gauge bosons.
}
As we will see below, these WZ terms can arise from integrating out
heavy anomaly-cancelling fermions~\cite{DHoker:1984izu,DHoker:1984mif}.
If the WZ terms are not to introduce additional electroweak symmetry
breaking --- for example, if the heavy fermions get their mass from a
SM-singlet vacuum expectation value --- then their coefficient
must be such as to cancel out the contribution of the $XBB$ and $XWW$
anomalies to the $W$ and $Z$ masses. Otherwise --- for example, if
the anomalies are cancelled by heavy SM-chiral fermions, which have been
integrated out --- the WZ coefficient may take other values.


\subsection{UV anomaly cancellation}
\label{sec:UV}
If the anomalies are cancelled by heavy fermions, then
for $U(1)_X$ to be preserved, the masses of these fermions
must be EW-breaking. In the simplest case, they could
obtain their masses through large Yukawa couplings with the
SM Higgs. 
As reviewed in Appendix~\ref{ap:chiral}, this possibility is strongly
constrained by electroweak precision tests and collider constraints;
assuming that the LHC run-II sees no deviations from the SM, it
will be fairly robustly ruled out.
An extended EWSB sector would alter the details, but is 
generically also subject to strong experimental constraints.

In UV completions where the heavy fermions have both SM-breaking
and $U(1)_X$-breaking masses, the WZ coefficient in the low-energy
theory depends on the relative size of these contributions. If the SM-breaking contributions are small compared to the total
masses, then the WZ coefficient will be approximately that expected from
a SM-preserving theory, up to $(m_{\rm EW} / m_f)^2$ corrections.
Conversely, if the dominant contribution to the masses is from a
EW-breaking VEV, then the situation will be approximately that
of the previous paragraph, and similar experimental constraints will
apply.

A caveat to bear in mind is that such constraints rely on the existence
of new, SM-chiral fermions, which have effects (such as electroweak
precision observables) unsuppressed by the small coupling $g_X$.
Within the low-energy theory, the effects of the SM-breaking WZ terms
are all suppressed by $g_X$, and if this is small enough, may not be
problematic. Consequently, it is possible that there exist more exotic
$U(1)_X$-preserving UV completions, without anomaly-cancelling fermions,
that are experimentally viable.

For the rest of this paper, we will focus on UV completions
which result in a SM-preserving effective theory.
These are easily realised; it is always possible to introduce
a new set of fermions with vectorial couplings to the SM gauge
bosons, but chiral couplings to $X$, along with a $U(1)_X$-breaking
VEV, to cancel the anomalies~\cite{Batra:2005rh}.
As noted above, the lack of new EWSB
fixes the coefficient of the WZ terms in the low-energy theory.
While the value of the coefficient depends on the regularisation scheme
chosen, the physical results are of course scheme-independent
(see Appendix~\ref{ap:anom}).
Since the EFT breaks $U(1)_X$, these results include energy-enhanced emission
of the longitudinal $X$ component, as we show below.

For UV completions with heavy anomaly-cancelling fermions,
a slight complication is that the new `UV' degrees of freedom do not necessarily
have to be heavier than all of the SM states. 
For example, in the case of a vector coupled to the SM baryon number current,
if we assume that anomalies are cancelled by SM-vector-like fermions, then collider constraints require that they
have masses $\gtrsim 90 \GeV$~\cite{Dobrescu:2014fca}.
If they are only slightly heavier than this bound,
then for external momenta around the scale of the SM EW boson masses,
the mass of the fermions in the loop will have an effect on anomalous
$X_L$ amplitudes. In the following, we will assume that
new states contributing to the anomaly are heavy enough that
such momentum dependence can ignored, except where otherwise stated.
This assumption will not be consistent for small enough $m_X/g_X$, since
within the SM + $X$ EFT, the growth of amplitudes with energy (as derived
below) requires that there are new states at a scale $\lesssim \frac{4 \pi m_X}{g_X}
/\left(\frac{3 g^2}{16 \pi^2}\right)$~\cite{Preskill:1990fr}. However, such large $g_X$
will generally be constrained more directly.


\subsection{Triangle diagram amplitudes}
\label{sec:xlamps}

To illustrate how the results outlined above arise,
we will compute anomalous triangle amplitudes within the low-energy
theory, and then show how this relates to the calculation
in a UV-complete theory.
Using the regularisation scheme from Appendix~\ref{ap:anom}
that is symmetric between external legs,
the longitudinal $XBB$ triangle amplitude is, summing over the SM
fermions in the loop,
\begin{gather}
	-(p + q)_\mu \MM^{\mu\nu\rho}_{\rm SM} = \frac{{\cal A} _{ XBB}}{12\pi^2} g_X g'^2 \epsilon^{\nu\rho\lambda\sigma} p_\lambda q_\sigma\,, 	\label{eq:smbanom} \\[0.5ex] 
	\hspace{0.5cm}	{\cal M} ^{\mu\nu\rho}_{\rm SM} \equiv 
	\sum_f\hspace{-0.2cm}\adjustbox{valign=m}{
 \begin{tikzpicture}[line width=0.75] 
\coordinate (C1) at (.75,0);
\coordinate (C2) at (0.75+0.75,{0.75*0.7});
\coordinate (C3) at (0.75+0.75,-{0.75*0.7});
\coordinate (C4) at (0.75+0.75+0.75,{0.75*0.7});
\coordinate (C5) at (0.75+0.75+0.75,-{0.75*0.7});
    \draw[v] (0,0) node[left]{$ X _\mu  $} -- (C1);
    \draw[f] (C1) -- (C2);
    \draw[f] (C2) -- (C3) node[midway,right] {$ f $};
    \draw[fb] (C1) -- (C3);
    \draw[v] (C2) -- (C4) node[right] {$ B _\nu  $} node[above,midway] {$  p \rightarrow  $};
    \draw[v] (C3) -- (C5)node[right] {$ B _\rho  $}node[below,midway] {$  q \rightarrow  $}; 
  \end{tikzpicture}} \notag \,.
\end{gather}
As reviewed in Appendix~\ref{ap:anom}, since the SM fermions have
vectorial couplings to $X$, this amplitude does not depend on the masses
of the SM fermions. 
This means that the momentum dependence of the longitudinal $X$ amplitude has the
simple $\epsilon^{\nu\rho\lambda\sigma} p_\lambda q_\sigma$
form, rather than involving extra terms depending
on the external momenta compared to the mass of the fermions in the loop.

The regularisation scheme being symmetric between external legs
means that we also have longitudinal $B$ amplitudes, $p_\nu
{\cal M} ^{\mu\nu\rho}_{\rm SM} = \frac{{\cal A} _{XBB}}{12\pi^2} g_X g'^2
\epsilon^{\mu\rho\lambda\sigma} q_\lambda p_\sigma$ etc (ignoring
the SM fermion masses). To get rid of
these, and restore the SM gauge symmetry within the SM + $X$ EFT, 
we need an explicit Wess-Zumino term,
\begin{equation}
	\LL \supset \frac{\Am_{XBB}}{12 \pi^2} g_X g'^2 \epsilon^{\mu\nu\rho\sigma} X_\mu B_\nu \partial_\rho B_\sigma  \,,
	\label{eq:wz1}
\end{equation}
that gives a contribution to the amplitude of
\begin{align}
   -(p + q)_\mu \MM_{\rm WZ}^{\mu\nu\rho} & = \frac{{\cal A} _{ XBB}}{6\pi^2} g_X g'^2 \epsilon^{\nu\rho\lambda\sigma} p_\lambda q_\sigma\\ 
  p _\nu {\cal M} _{ \rm WZ} ^{ \mu\nu\rho } & = - \frac{ {\cal A} _{ XBB} }{ 12 \pi ^2 } g _X g ^{ \prime 2 } \epsilon ^{ \mu \rho \lambda \sigma } q _\lambda p _\sigma \\
  q _\rho   {\cal M} _{ \rm WZ} ^{ \mu\nu\rho } & = - \frac{ {\cal A} _{ XBB} }{ 12 \pi ^2 } g _X g ^{ \prime 2 } \epsilon ^{ \mu \nu  \lambda \sigma } q _\lambda p _\sigma\,.
\end{align}
Adding together the contributions from the WZ term and from the SM
fermion triangle diagrams, we obtain a total
amplitude $\MM \equiv \MM_{\rm SM} + \MM_{\rm WZ}$ of
\begin{gather}
-(p + q)_\mu \MM^{\mu\nu\rho} = \frac{{\cal A} _{ XBB}}{4\pi^2} g_X g'^2 \epsilon^{\nu\rho\lambda\sigma} p_\lambda q_\sigma  \notag   \\ 
p_\nu \MM^{\mu\nu\rho} = q_\rho \MM^{\mu\nu\rho} = 0  \,,\label{eq:beft1}  \\[0.5ex]
 	{\cal M} ^{\mu\nu\rho} \equiv 
\sum_f\hspace{-0.2cm}\adjustbox{valign=m}{
 \begin{tikzpicture}[line width=0.75] 
\coordinate (C1) at (.75,0);
\coordinate (C2) at (0.75+0.75,{0.75*0.7});
\coordinate (C3) at (0.75+0.75,-{0.75*0.7});
\coordinate (C4) at (0.75+0.75+0.75,{0.75*0.7});
\coordinate (C5) at (0.75+0.75+0.75,-{0.75*0.7});
    \draw[v] (0,0) node[left]{$ X _\mu  $} -- (C1);
    \draw[f] (C1) -- (C2);
    \draw[f] (C2) -- (C3) node[midway,right] {$ f $};
    \draw[fb] (C1) -- (C3);
    \draw[v] (C2) -- (C4) node[right] {$ B _\nu  $} node[above,midway] {$  p \rightarrow  $};
    \draw[v] (C3) -- (C5)node[right] {$ B _\rho  $}node[below,midway] {$  q \rightarrow  $}; 
  \end{tikzpicture}} + 
\adjustbox{valign=m}{ \begin{tikzpicture}[line width=0.75] 
\coordinate (C1) at (0,0);
\coordinate (C2) at (0.75,0);
\coordinate (C3) at (0.75+0.75,{0.75*0.7});
\coordinate (C4) at (0.75+0.75,-{0.75*0.7});
    \draw[v] (C1)node[left] {$ X $} -- (C2) ;
    \draw[v] (C2) -- (C3) node[right] {$ B $};
    \draw[v] (C2) -- (C4)node[right] {$ B $}; 
    \draw[fill=black] (C2) circle (0.1);
  \end{tikzpicture}}
	\,\notag 
\end{gather}
with the SM gauge symmetry now preserved. 

The motivation for adopting a symmetric regularisation scheme is
that it makes clear how this amplitude, calculated within the SM + $X$ EFT,
relates to the calculation within a 
UV theory. The simplest
UV completion, as discussed above, cancels the anomalies by introducing
extra fermions which couple vectorially to the SM gauge bosons, but
axially to $X$. These obtain heavy masses from a $U(1)_X$-breaking,
but SM-singlet, VEV.
In this setup, the `anomalous' contributions to the $XBB$ amplitude
cancel between the new fermions and the SM fermions (in any regularisation
scheme),
leaving only the fermion-mass-dependent pieces. 
Since the SM
fermions have vectorial couplings to $X$, the mass dependence
is only on the new fermions, which have axial $X$ couplings.
The total longitudinal amplitude is
	\begin{align}
		 	-(p + q)_\mu \MM^{\mu\nu\rho} &=  \frac{1}{2\pi^2}\epsilon^{\nu\rho\lambda\sigma} p_\lambda q_\sigma 
	g_X g'^2 \times
		\label{eq:triuv}
		\\
		&\sum_{f}  2 m_{f}^2 I_{00}(m_{f},p,q) X_{A,f} Y_{f}^2\,,
		\notag
	\end{align}
where $f$ runs over the new fermions, $Y_f$ is the hypercharge of $f$,
$X_{A,f}$ is the axial coupling of $f$ to $X$,
and the mass-dependent `scalar integral'~\cite{AdlerAnomalies} term
is 
	\begin{align}
& 	I_{00}(m_f,p,q)	\equiv \int_0^1 dx \int_0^{1-x} dy\, \frac{1}{D ( x, y ,p,q) }\,, \\ 
& D \equiv y(1-y) p^2 + x(1-x) q^2 	+ 2 x y \, p \cdot q - m_f^2 \,.\notag
	\end{align}
For fermion masses well below the external momenta, i.e., $ m_f^2 \ll
p^2, q^2, p \cdot q $, the mass-dependent term vanishes, $ 2 m _f ^2 I
_{ 00} \simeq 0 $. This indicates that if there is no mass gap between
the SM and heavy fermions, the longitudinal $X$ amplitude cancels between
the two sectors.
At the other extreme, if $ m_f^2 \gg p^2, q^2, p \cdot
q $ the mass dependent term approaches a constant, independent of heavy
fermion mass, $ 2 m _f ^2 I _{ 0 0} \simeq - 1 $.
The key here is that anomaly cancellation in the UV requires that
\begin{equation}
 \Am_{XBB} = 
	-2 \sum_f X_{A,f} Y_f^2
\end{equation}
(the factor of 2 on the right-hand side comes that the fact that $f$ runs over the heavy
Dirac fermions, each of which is made up of two Weyl fermions).
 Hence, if the masses of the heavy fermions are much greater than the external momenta then the total
amplitude in equation~\ref{eq:triuv} is
\begin{equation}
	-(p + q)_\mu \MM^{\mu\nu\rho} \simeq  \frac{{\cal A} _{ XBB}}{4\pi^2}\epsilon^{\nu\rho\lambda\sigma} p_\lambda q_\sigma 
	g_X g'^2 
		 \,,
\end{equation}
giving the same result as the EFT calculation (equation~\ref{eq:beft1}).
We emphasize that this result is independent of the details of the UV theory,
and only depends on it not introducing extra SM breaking.
Equation~\ref{eq:triuv} also illustrates
how integrating out the heavy fermions gives the WZ term from
equation~\ref{eq:wz1}, and how there will be $\sim p/m_f$ etc.\
corrections, corresponding to higher-dimensional operators within the
effective theory.

The amplitudes for $XWW$ triangles will have similar behaviour,
with $g'$ replaced by $g$. An additional feature is that, since
$SU(2)_L$ is non-abelian, there are anomalous $XWWW$ box diagrams,
corresponding to the $XWWW$ part of the WZ term in equation~\ref{eq:WZ}.
These have an analogous story of fermion mass dependence in the UV theory.

By introducing gauge degrees of freedom for the longitudinal modes of the vector
bosons,
we could equivalently have calculated triangle
diagrams between SM gauge bosons and the Goldstone mode for $X$. 
The latter has couplings $\propto m_f$ to the heavy fermions,
so their contribution to triangle amplitudes becomes
$\sim$ constant in the heavy $m_f$ limit (in exact analogy to
fermions with large Yukawa couplings in the SM, as discussed
in~\cite{DHoker:1984izu,DHoker:1984mif}).


\subsection{Axion-like behaviour}

By the usual Goldstone boson equivalence arguments,
the $1/m_X$-enhanced parts of amplitudes involving
longitudinal $X$ are $\simeq$ to those for the corresponding
Goldstone (pseudo)scalar, $\varphi$. Stated
more precisely, we can take the limit 
$m_X \rightarrow 0$, $g_X
\rightarrow 0$ with $f_X \equiv m_X/g_X$ held constant,
decoupling the transverse $X$ modes.
Then, making the substitution $g_X X_\mu \mapsto \frac{1}{f_X}
\partial_\mu \varphi$ gives the same amplitudes
in the $X$ and $\varphi$ theories.
For finite $m_X$, they will be equal up to $\OO(m_X/E)$, where
$E$ is some scale associated with the process.\footnote{
	Processes involving large energies in a particular
	frame may still involve small invariant energy scales
	for the anomalous couplings. For example, if $X$ has an anomalous
	coupling to photons, then the production of an on-shell $X$ from two high-energy on-shell photons involves energies $\sim m_X$ in the rest frame of the $X$,
	and is in fact forbidden by the Landau-Yang theorem. In such cases,
	the $X$ amplitude can be $\OO(1)$ different from Goldstone
	one.
	}

In our case, the $X_L$ processes which survive in this limit
all come from the anomalous couplings. 
In the $\varphi$ theory, we can integrate by parts
to write these couplings as
		\begin{align}
			&\frac{\Am_{XBB}}{16 \pi^2} \frac{\varphi}{f_X} (g^2 W^a \tilde{W}^a - g'^2 B \tilde{B})
			= \notag \\
			&\frac{\Am_{XBB}}{16 \pi^2} \frac{\varphi}{f_X} \left(
			g^2 (W^+ \tilde{W}^- + W^- \tilde{W}^+) \right. \notag \\
			&\quad \left. + g g' (\cot\theta_w - \tan\theta_w) Z \tilde{Z} + 2 g g' Z \tilde{F}) \right.
			\notag
			\\
			&\quad \left. - i e g^2 \tilde{F}^{\mu\nu} (W^+_\mu W^-_\nu -
			W^+_\nu W^-_\mu) + \dots \right)\,,
			\label{eq:bphic}
		\end{align}
		where we have suppressed indices, and
the dots correspond to further terms of the form $AW^+ W^-$
and $Z W^+ W^-$.\footnote{the $WWWW$ terms from $W^a_{\mu\nu} (\tilde{W}^a)^{\mu\nu}$
cancel, reflecting the lack of pentagon anomalies for an abelian
vector~\cite{Bilal:2008qx}.}
Thus, energy-enhanced $X_L$ emission processes will have the same leading
rate as the emission of an axion-like-particle (ALP) with these SM gauge
boson couplings. This means that we can use the same processes that are
used to search for light ALPs to look for $X$. 

As mentioned above, 
within the simplest kinds of UV-complete models, 
the origin of equation (\ref{eq:bphic})
can be traced to
explicit Yukawa interactions of anomaly-cancelling fermions with a set of $U(1)_X$ complex Higgs fields, 
for which $\varphi$ is the Goldstone mode.
Integrating out these fermions gives the couplings in 
equation (\ref{eq:bphic}), in exact analogy with the axion literature.

Since there is no two-photon anomalous coupling (as $X$ has vectorial
couplings to the SM fermions), longitudinal emission processes involving
sub-EW-scale momenta are suppressed. Consequently, the
most important effects of the anomalous couplings arise
either in high-energy collisions --- for example, on-shell $Z$
decays --- or in virtual processes which can be dominated
by large loop momenta, such as rare meson decays.



\subsection{$Z \rightarrow \gamma X$}
\label{sec:zgx}

If $m_X < m_Z$, then the $\varphi Z \tilde{F}$ coupling 
in equation~\ref{eq:bphic} gives rise to $Z \rightarrow
\gamma X$ decays, with width
\begin{align}
\label{GaZ}
	\Gamma_{Z \rightarrow \gamma X_L} &\simeq 
	\frac{|\Am_{XBB}|^2}{1536 \pi^5} g_X^2 g^2 g'^2 \frac{m_Z^3}{m_X^2} \,,\\
	&= 
	\frac{|\Am_{XBB}|^2}{1536 \pi^5} g^2 g'^2 \frac{m_Z^3}{f_X^2} \,,\notag
\end{align}
corresponding to a branching ratio
\begin{equation}
	\frac{\Gamma_{Z \rightarrow \gamma X_L}}{\Gamma_Z} \simeq
	3 \times 10^{-8} |\Am_{XBB}|^2 \left(\frac{\TeV}{f_X}\right)^2\,.
\end{equation}

The corresponding experimental signatures and constraints
depend on how $X$ decays.
At small $m_X$ and $g_X$, the $X$ decay length will be longer
than the scale of the experiment, so will give a missing 
energy signature
(at the small $g_X$ we are interested in, $X$ will generally not
interact strongly enough to be detected by its scattering).
If $X$ escapes the detector or decays invisibly,  then
LEP searches~\cite{Acciarri:1997im,Abreu:1996pa} for single photons
at half the $Z$ energy constrain this branching ratio to be $\lesssim
10^{-6}$.

For visible decays, there are published branching ratio limits
for $ X \rightarrow $ jet + jet and $ X \rightarrow l ^+ l ^- $ of $
\lesssim 3 \times 10 ^{ - 3 } $~\cite{Adriani:1992zm} and $ 5 \times 10
^{ - 4 } $~\cite{Acton:1991dq} respectively, with some improvement in
the high mass region~\cite{Adeva:1991dw}. 
While these limits (from LEP) are 
not particularly stringent, we expect that the LHC has the capacity
to significantly improve them.
As illustrated by searches such as~\cite{Aaltonen:2013mfa,Khachatryan:2015rja},
it is likely that leptonic $Z \rightarrow \gamma (X \rightarrow l^+ l^-)$
decays could be probed down to $\OO(10^{-5})$ branching
ratios or better.

Enhanced $X_L$ emission also occurs for processes involving off-shell EW
gauge bosons, with higher-energy processes giving increasing enhancements
up to the scale of new states. 
For example, the rate for high-energy annihilation of SM fermions,  $\psi \bar\psi \to \gamma^* \to ZX$,
will be enhanced by $(E/m_X)^2$, where $E$ is the center of mass energy of the colliding particles. 
In some models, off-shell anomalous production
may be the most promising LHC search channel, e.g.\ when $X$ decays invisibly
or escapes from the detector.


\subsection{FCNCs}
\label{sec:anomfcnc}

The coupling of $X$ to quarks and the anomalous $XWW$ coupling both
lead to flavor changing neutral current (FCNC) interactions.
Since the most important effects of these are at meson energy scales,
the simplest procedure is to integrate out EW-scale states
to obtain an effective $X q q'$ vertex. The QCD coupling
is small at scales $\sim m_W$, so the calculation is under
perturbative control (see e.g.~\cite{Buchalla:1992zm}). 
The leading effective interaction introduced is
\begin{gather}
	\LL \supset g_{X d_i d_j} X_\mu \bar{d}_j \gamma^\mu \mathcal{P}_L d_i + {\rm h.c.} + \dots \,,\label{eq:lgx} \\
	\adjustbox{valign=m}{
	\begin{tikzpicture}[line width=1] 
    \begin{scope}[shift={(0,0)}]
      \draw[f] (0,0) node[below] {$ d _i  $}-- (1,0);
          \draw[decoration={markings,mark=at position 0.7 with {\arrow[very thick]{latex}}},postaction={decorate}] (1,0) -- (2,0) node[below] {$ d _j  $};
          \draw[v] (1,0) -- (2,0.75) node[left,xshift=-0.2cm] {$ X $};
          \draw[pattern=north west lines,preaction={fill=white}] (1,0) circle (0.2);
        \end{scope}
          \begin{scope}[shift={(2.4,0)}]
            \draw[f] (0.33,0) 
            -- (1.2,0);
			\draw[f] (1,0) -- (2,0) node[above,xshift=-0.5cm] {$u/c/t$};
            \draw[f] (2,0) -- (2.66,0);
            \draw[v] (1,0) arc (180:360:0.5);
            \node[] at (.8,-0.6) {$W$};
            \draw[fill=black] (1.5,-0.52) circle (0.13);
            \draw[v] (1.5,-0.5) -- (2.25,-1.25);
          \end{scope}
          \begin{scope}[shift={(5,0)}]
            \draw[f] (0.33,0) --(1.2,0);
            \draw[f] (1,0) -- (2,0);
            \draw[f] (2,0) -- (2.66,0);
            \draw[v] (1,0) arc (180:360:0.5);
            \draw[v] (1.5,0) -- (2,0.75) ;
            \node[] at (1.5,-0.9) {$W$};
          \end{scope}
          \node[] at (2.35,0.25) {\large $  \bm{=}$};
          \node[] at (5.2,0.25) {\large $  \bm{+}$};
	\end{tikzpicture}} \notag\,,
\end{gather}
where we have taken a down-type FCNC for illustration, and have
omitted other, higher-loop-order diagrams (as well as $X$ emission
from external quark legs). The solid $XWW$ vertex indicates the sum
of WZ terms and fermion triangles (within a UV theory, it would simply
be the sum over triangles). If $X$ is coupled to a fully-conserved
current, then $g_{Xd_i d_j} = 0$, and the effective interaction is
higher-dimensional;
if $X$ is coupled to a tree-level conserved current (as we consider here), then only the anomalous
$XWW$ coupling contributes to $g_{X d_i d_j}$.

This effective operator then gives rise to flavor-changing meson
decays. For heavy-quark decays, such as $b \rightarrow
s X$, these rates depend on hadronic
matrix elements, which can be derived from QCD
light-cone sum rules~\cite{Ball:2004ye,Ball:2004rg}.
For kaon decays, we can use chiral perturbation theory
to obtain the leading approximation to the decay rates.
Since the renormalisation of the left-handed quark current
is proportional to the quark masses, the RG evolution of the effective
operator from scales $\sim m_W$ to meson energy scales can sensibly
be ignored, in a first approximation.
This interaction leads to meson decay rates through $X_L$ emission of
\begin{align}
	&\Gamma(B \rightarrow K X) \simeq \frac{m_B^3}{64 \pi m_X^2} |g_{bsX}|^2 
	\left(1 - \frac{m_{K}^2}{m_{B}^2}\right)^2
	|f_K(m_X^2)|^2 \frac{2 Q}{m_B}\,, \\
	&\Gamma(B \rightarrow K^* X) \simeq \frac{m_B^3}{64 \pi m_X^2} |g_{bsX}|^2 
	|f_{K^*}(m_X^2)|^2 \left(\frac{2 Q}{m_B}\right)^3 \,,\\
	&\Gamma(K^\pm \rightarrow \pi^\pm X) \simeq \frac{m_{K^\pm}^3}{64 \pi m_X^2}
	\left(1 - \frac{m_{\pi^\pm}^2}{m_{K^\pm}^2}\right)^2|g_{sdX}|^2 \frac{2 Q}{m_{K^\pm}}\,,
	\\
	&\Gamma(K_L \rightarrow \pi^0 X) \simeq \frac{m_{K_L}^3}{64 \pi m_X^2}
	\left(1 - \frac{m_{\pi^0}^2}{m_{K_L}^2}\right)^2{\rm Im}(g_{sdX})^2 \frac{2 Q}{m_{K_L}}\,,
\end{align}
where $Q$ is the momentum of the decay products in the center of mass frame, and
$f_K, f_{K^*}$ are the appropriate form factors coming from the hadronic matrix
elements (we use the fits from~\cite{Ball:2004ye,Ball:2004rg}).
The deviation of the kaon decay rate from the
leading chiral perturbation theory value given 
above is expected to be of order a few percent
(see e.g.~\cite{Mescia:2007kn}). 
Since we are quoting the leading $1/m_X^2$ rates, 
we would obtain same-order results by evaluating the
form factors at zero momentum transfer, $f_{K}(m_X^2) \simeq f_{K}(0)$ etc. 

In the calculation of $g_{d_i d_j X}$, while each individual diagram in \eqref{eq:lgx} is
divergent, these divergences
cancel in the sum over virtual up-type quarks.
This occurs since the divergent terms are independent of the quark
mass, so their sum cancels due to the unitarity of the CKM matrix.
As a result, the integral is dominated by momenta $\sim m_t$,
and couplings suppressed by the cutoff scale will give sub-leading
contributions (in the UV theory, the masses of the UV fermions
in triangles will be much larger than the external momenta of these triangles).
The coefficient of the effective vertex is
\begin{equation}
	g_{X d_i d_j} = -\frac{3 g^4 \Am_{XBB}}{16 \pi^2} g_X
	\sum_{\alpha \in \{u,c,t\}} V_{\alpha i} V_{\alpha j}^* F\left(\frac{m_\alpha^2}
	{m_W^2}\right) + \dots\,,
	\label{eq:2loop}
\end{equation}
where
\begin{equation}
	F(x) \equiv \frac{x(1 + x (\log x - 1))}{(1 - x)^2}
	= x + \OO(x^2 \log x)\,.
\end{equation}
Due to the $m_q^2/m_W^2$ dependence for small quark mass, the sum over
up-type quarks is dominated by the top quark, for both $bsX$ and $sdX$
vertices, despite the smallness of the $V_{ts}$ and $V_{td}$ CKM
matrix elements.
Since $m_b^2/m_W^2$ is small, the equivalent up-type FCNC
vertices, such as $cuX$, are suppressed compared to down-type 
FCNCs.

Compared to these effective FCNC vertices, other effective flavor-changing
operators are higher-dimensional, and so are suppressed 
by more powers of $g_X/m_X$ and/or $1/m_W^2$.
Thus, despite equation~\ref{eq:2loop} representing a 2-loop contribution
(within the UV theory), it is able to dominate over 1-loop $d_i d_j X$ processes.
For example, in the $B \rightarrow K X$ decay we have,
\begin{equation}
	\MM^{{\rm 2-loop}} / \MM^{{\rm 1-loop}} \propto g^2/(16 \pi^2)
	\times (m_t / m_X)^2\,,
	\label{eq:twor}
\end{equation}
which, for $m_X$ light enough to be emitted in the decay, is $\gg 1$.\footnote{
	The $\propto m_X^2$ (rather than $\propto m_X$) relative suppression
	of 1-loop emission comes from 
angular
momentum conservation in the pseudoscalar $\rightarrow$ pseudoscalar + vector decay;
for $B \rightarrow K^* X$ decays, we would have $\MM^{{\rm 2-loop}} / \MM^{{\rm 1-loop}} \propto m_t^2/(m_X m_b)$ instead.}
Competing SM FCNC processes are also suppressed; for example,
 the $bs \gamma$
vertex is of the form $\propto\frac{m_b}{m_W^2} F_{\mu\nu} \bar{b}_L
\sigma^{\mu\nu} s_L + \dots $ \cite{Inami:1980fz,Misiak:2006zs} (on-shell), since the photon 
couples to a conserved current, while 4-fermion vertices are suppressed by
at least $G_F$.

If $m_X$ is light enough, then FCNC meson decays via an on-shell
longitudinal $X$ become possible, and are enhanced by $({\rm energy} /
m_X)^2$, in addition to being lower-dimensional than other
effective flavor-changing processes. Most directly, the $bsX$ and $sdX$ vertices
result in $B \rightarrow K^{(*)} X$ and
$K \rightarrow \pi X$ decays, 
giving new flavor-changing meson decays
that can place strong constraints on the coupling of $X$.
This is in exact analogy to the 
FCNC processes discussed in~\cite{Izaguirre:2016dfi}, for axion-like particles with a coupling to $W^a \tilde{W}^a$.
In contrast, processes involving two or more $d_i d_j X$ vertices,
such as the $X$ contribution to meson oscillations, 
are suppressed by $1/f_X^2$, but compete with SM processes
suppressed by $1/m_W^2$. Consequently, it is difficult for such
processes to probe $f_X$ above the EW scale, unless $ m _X $ is accidentally very close to the meson mass, resulting in resonant enhancement from meson-$ X $ mixing.

The selection rules for decays via longitudinal vector emission
are different to those for transverse emission.
In the latter case,
angular momentum conservation suppresses (pseudo)scalar $\rightarrow$
(pseudo)scalar + vector decays, since these demand that the vector's spin
is perpendicular to its momentum.
This suppresses the rate of such decays via a vector that couples to
a conserved current; for example, 
there are no $B^+ \rightarrow K^+ \gamma$ decays,
while the rate for $B^+ \rightarrow K^+ A'$, where $A'$ is a kinetically-mixed dark photon,
is proportional to $m_{A'}^2$ \cite{Pospelov:2008zw}.
However, by Goldstone boson equivalence, meson decays via a light longitudinal
$X$ have the same rates as the corresponding ALP decays, so
decays such as $B^+ \rightarrow K^+ X$ are unsuppressed.

\subsubsection{Experimental constraints}
\label{sec:fcncexp}

Here, we summarise the experimental searches we will use to constrain
FCNC meson decays via $X$ --- for easy reference, these are tabulated
in Table~\ref{tab:FCNC}.  
\begin{table*}
\center
\bgroup
\def\arraystretch{1.5}
\begin{tabular}{lcccc}
\toprule[1.5pt]
Decay type & Measured branching ratio & Reference  & Comments  \\ 
\midrule[1pt] 
$ B   \rightarrow K   ( X \rightarrow \ell  ^+ \ell  ^- )   $ &$ ( 4.7 \pm 0.6 \pm 0.2 )  \times 10 ^{ - 7} $ & BaBar, 2012~\cite{Lees:2012tva}  & $ m _{ \ell \ell  } ^2  >0.1~\text{GeV} ^2 $ \\ 
$ B   \rightarrow K (   X  \rightarrow \text{inv} )   $ &$ < 2.5   \times 10 ^{ - 5} $ & Belle, 2017~\cite{Grygier:2017tzo}  & \\ 
$ B ^+ \rightarrow K ^+ ( X \rightarrow \mu  ^+ \mu  ^-  ) $  &$ ( 4.36 \pm 0.15 \pm 0.18 ) \times 10 ^{ - 7} $ & LHCb, 2012 ~\cite{Aaij:2012vr} & $ m _{ \mu \mu   } ^2  >0.05~\text{GeV} ^2 $\\ 
  $ B ^0 \rightarrow K  ( X \rightarrow 3 \pi )    $  & $ < 2.3 \times 10 ^{ - 4 } $ & Particle Data Group~\cite{Olive:2016xmw} &  \\
  $ B ^0 \rightarrow K  ( X \rightarrow \mu ^+ \mu ^-  )    $  & $< 2 \times 10 ^{ - 10} $ -- $  10 ^{ - 7 }  $& LHCb, 2016~\cite{Aaij:2016qsm} & displaced search \\          \midrule[0.5pt]
$ B \rightarrow K ^{ \ast } ( X   \rightarrow \ell  ^+ \ell  ^- )  $ & $ ( 10.2 ^{ + 1.4} _{ -1.3} \pm 0.5  )  \times 10 ^{ - 7} $& BaBar, 2012~\cite{Lees:2012tva}  & $ m _{ \ell \ell  } ^2  > 0.1~\text{GeV}^2  $ \\ 
$ B   \rightarrow K ^\ast  ( X \rightarrow \text{inv} )   $ &$ < 1.6   \times 10 ^{ - 5} $& Belle, 2017  ~\cite{Grygier:2017tzo}  & \\ 
$ B ^+ \rightarrow K ^{\ast +} ( X \rightarrow e  ^+ e  ^- )  $  &$ ( 1.32 ^{ + 0.41 } _{ - 0.36} \pm 0.09 ) \times 10 ^{ - 6} $ & BaBar, 2008~\cite{Aubert:2008ps} & $ m _{ e e    } ^2   < 0.1 \text{ GeV} $ \\ 
$ B ^0 \rightarrow K ^{ \ast 0} ( X \rightarrow e  ^+ e  ^-  ) $  &$ ( 0.73 ^{ + 0.22 } _{ - 0.19} \pm 0.04 ) \times 10 ^{ - 6} $ & BaBar, 2008~\cite{Aubert:2008ps} &  $ m _{ e e    } ^2   < 0.1 \text{ GeV} $  \\ 
\midrule[0.5pt]
$ K _L \rightarrow \pi ^0    ( X \rightarrow e  ^+ e  ^- )   $ & $ < 2.8 \times 10 ^{ -10} $ & KTeV/E799, 2003~\cite{AlaviHarati:2003mr}  &  $ m _{ ee} > 140 \text{ MeV} $\\ 
$ K _L \rightarrow \pi ^0    ( X \rightarrow \mu   ^+ \mu   ^- )  $ & $ < 3.8 \times 10 ^{ -10} $ & KTeV, 2000~\cite{AlaviHarati:2000hs}  &  \\ 
$ K ^\pm  \rightarrow \pi ^{ \mp }    ( X \rightarrow \mu   ^+ \mu   ^- )   $ & $ < 1.1 \times 10 ^{ -9}  $ & NA48/2, 2011 ~\cite{Batley:2011zz}  &  \\
$ K ^+  \rightarrow \pi ^{ + }    ( X  \rightarrow \text{inv} )  $ & $ (1.73 ^{ + 1.15 } _{ - 1.05} ) \times 10 ^{ - 10} $ & E949,2008~\cite{Artamonov:2008qb}  &  \\
\bottomrule[1.5pt]
\end{tabular}
\egroup
\caption{Summary of the FCNC searches used to constrain models with a light vector $X$, in the text and in figures.}
\label{tab:FCNC}
\end{table*}

If $X$ is sufficiently light and weakly coupled that it decays outside
the detector, then $B \rightarrow K \nu\bar\nu$ and $K \rightarrow \pi
\nu\bar\nu$ searches constrain the $B \rightarrow K X$ and $K \rightarrow
\pi X$ branching ratios.
The $K \rightarrow \pi \nu\bar \nu$ channel is especially constraining,
with existing experiments having measured a very small ($\sim 10^{-10}$)
branching fraction consistent with the SM prediction~\cite{Anisimovsky:2004hr,Artamonov:2008qb},
which the future NA62 experiment should be able to
measure to $\sim 10\%$ relative
error~\cite{Anelli:2005ju}.

For prompt decays of $X$ into leptons, as can occur
for heavier / more strongly coupled $X$, searches for $B \rightarrow K^{(*)} l^+ l^-$
and $K \rightarrow \pi l^+ l^-$ decays place strong constraints.
The LHCb search for $B^\pm \rightarrow K^\pm \mu^+ \mu^-$ decays
measures the branching ratio to be $(4.36 \pm 0.15 \pm 0.18) \times 10^{-7}$
\cite{Aaij:2012vr}. 
For kaons, the
$K^0_L \rightarrow \pi^0 e^+ e^-$ decay
is very well-constrained, 
with a branching ratio bound of $\lesssim 3 \times 10^{-10}$~\cite{AlaviHarati:2003mr}. However, because of the large hadronic branching
ratios for $K^0_L \rightarrow \pi^0 \pi^0$ and $K^0_L \rightarrow \pi^0 \pi^0 \pi^0$, the Dalitz decay $\pi^0 \rightarrow e^+ e^- \gamma$ gives
a background that makes $K^0_L \rightarrow \pi^0 e^+ e^-$ measurements
difficult at $m_{ee} \lesssim m_{\pi^0}$ \cite{AlaviHarati:2003mr} (the same applies
to $K^\pm \rightarrow \pi^\pm e^+ e^-$ versus $K^\pm \rightarrow \pi^\pm \pi^0$
\cite{Batley:2009aa}). Thus, for $m_X \lesssim m_{\pi^0}$,
the best constraints come from $B \rightarrow K^{(*)} e^+ e^-$ decays,
where the competing $B \rightarrow K \pi^0$ decays are also suppressed.
For example, the $B \rightarrow K^* e^+ e^-$
branching ratio is measured to be $\simeq 10^{-6}$ for $m_{ee} \lesssim
300 \MeV$~\cite{Aubert:2008ps}.

If $X$ dominantly decays into hadrons, then it may be possible
to perform bump-hunt searches in the invariant mass distributions
of $B \rightarrow K + {\rm hadronic}$ decays.
For example, the $B \rightarrow K \omega$ decay
is detected as a peak in the $m_{3 \pi}$ distribution
of $B \rightarrow K \pi^+ \pi^- \pi^0$ decays,
with branching ratio error $\sim 10^{-6}$~\cite{Chobanova:2013ddr};
a similar search could be performed at other invariant masses.

In addition to the prompt and invisible decays discussed above,
it is also possible to look for displaced $X$ decays.
LHCb performed a search for long-lived scalar particles decaying into $ \mu ^+ \mu ^- $ in $ B \rightarrow K X  $ decays~\cite{Aaij:2016qsm}. These limits dominate in the displaced regime with branching ratio limits reaching $ \simeq 2 \times 10 ^{ - 10 } $.
For very displaced decays, the best constraints come from beam
dump experiments. Here, the enhanced $K \rightarrow \pi X_L$ decay means
that kaon decays, which are usually a sub-dominant production
mechanism in proton beam dump experiments (for tree-level vector couplings),
can be the dominant process through which $X$s are produced.
This allows proton beam dump experiments such as 
CHARM~\cite{Bergsma:1985qz}\footnote{There has been some debate in the literature about the correct way to estimate the CHARM bounds, centered around neglecting Kaon absorption~\cite{Winkler:2018qyg}, the B meson energy~\cite{Dobrich:2018jyi}, and the geometric efficiency of particles hitting the detector \cite{Egana-Ugrinovic:2019wzj}. We take $ 2.4 \times 10 ^{ 18} $ protons-on-target, a Kaon absorption length of $15.3~{\rm cm}$, fraction of produced $ K ^+ $, $ K _L $, and $ B $ of 0.62,0.28, and $ 3.2 \times 10 ^{ - 7} $, respectively, $K$ and $ B $ momentum to be $ 25 ~{\rm GeV} $ and $ 75~{\rm GeV} $, respectively, geometric factors of $ X $ reaching the target for all meson decays to be 0.01, and all lepton reconstruction efficiencies to be 0.5.}  and (in the future) SHiP~\cite{Alekhin:2015byh} to probe smaller couplings
than indicated by a naive analysis~\cite{Alekhin:2015byh,Gardner:2015wea}.

It should be noted that, 
unlike constraints involving visible $X$ decays, missing energy searches
are effective down to arbitrarily small vector masses, and constrain 
correspondingly tiny $g_X$ for small $m_X$.
However, for $X$ with couplings to first-generation fermions, the strong
constraints coming from stellar energy loss bounds~\cite{Raffelt:1996wa,Hardy:2016kme},
and from 
fifth force / equivalence principle tests at smaller $m_X$~\cite{Graham:2015ifn}, mean that
it is generically only at extremely small $m_X$ that missing energy constraints
become the dominant bound.


\subsection{Baryon number coupled vector}
\label{sec:bvector}

To give an example of how these constraints relate to each other
and to other bounds in the literature, for a specific model,
we will consider a vector coupled to the SM baryon number
current. 
This model has been investigated in many papers over the past
decades, with 
motivations including
acting as a stabilisation mechanism for baryon number~\cite{Carone:1994aa},
mediating a new force that avoids the strong constraints coming
from leptonic, axial, or non-Minimal Flavor Violation couplings~\cite{Tulin:2014tya},
as well as addressing experimental anomalies~\cite{Feng:2016ysn}.

The SM baryon number current is conserved at tree level,
but broken by hypercharge and $SU(2)_L$ anomalies,
with $\Am_{XBB} = - \Am_{XWW} = n_g/2$, where $n_g = 3$
is the number of SM generations.  Consequently, subject
to caveats regarding the UV completion (see below),
there will be anomalous production processes of the kind
we have considered above.

Assuming that $X$ does not decay to hidden sector states,
then for $m_X \gtrsim m_\pi$, the vector will dominantly decay
to hadronic states (see~\cite{Tulin:2014tya} for a more detailed
analysis of decay channels). While a `pure' baryon-number-coupled
vector does not have tree-level couplings to leptons,
a kinetic mixing with the photon
is generated by RG evolution, so can
only be set to zero at a single scale. For commonality with other
literature, such as~\cite{Tulin:2014tya}, we will assume a kinetic mixing
$\epsilon = e g_X / (4 \pi)^2$ for our plots. Consequently, for $2m_e
< m_X \lesssim m_\pi$, $X$ will decay to $e^+ e^-$ through the
kinetic mixing.

\begin{figure*}
	\centering
		\includegraphics[width=.47\textwidth]{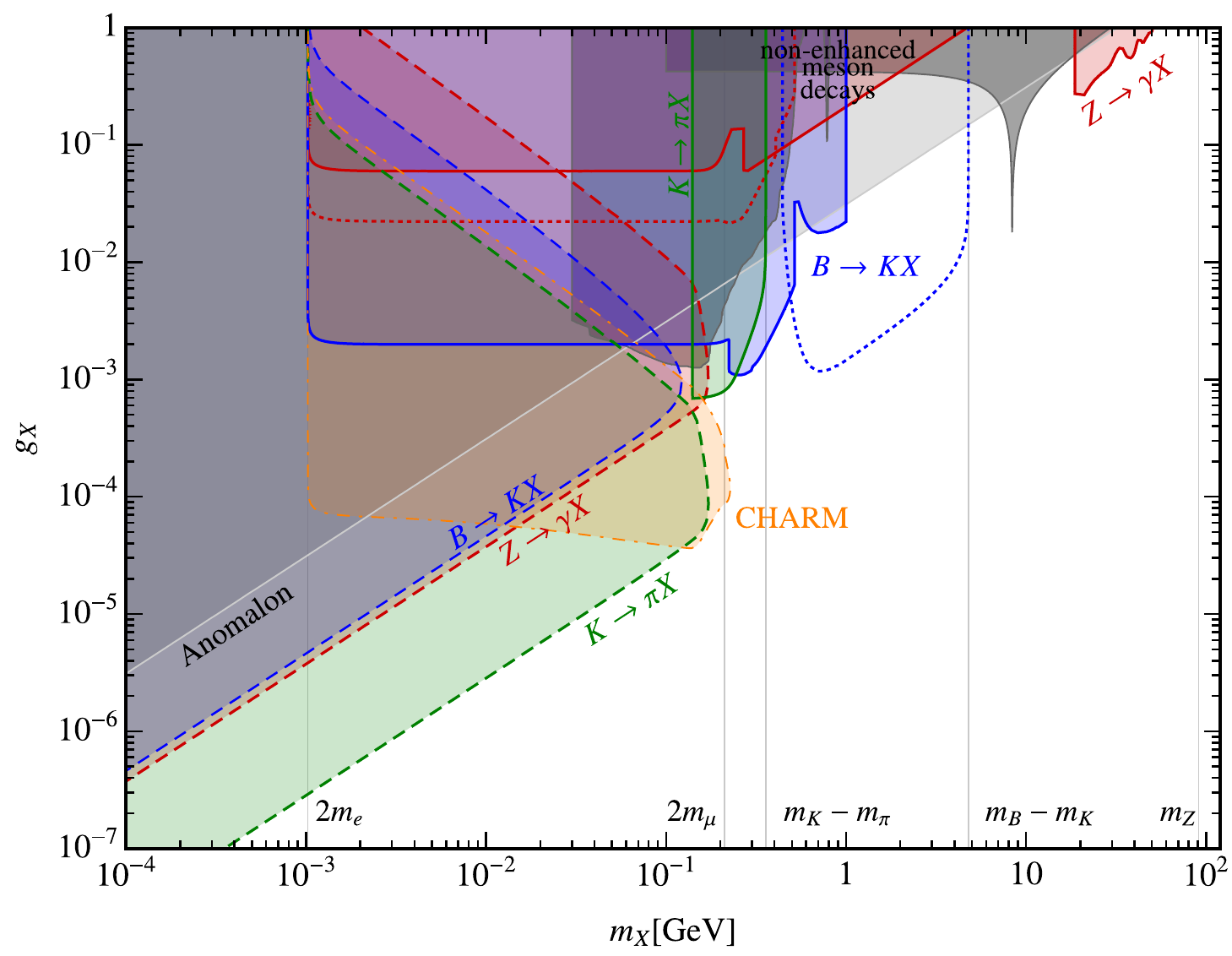}
		\includegraphics[width=.47\textwidth]{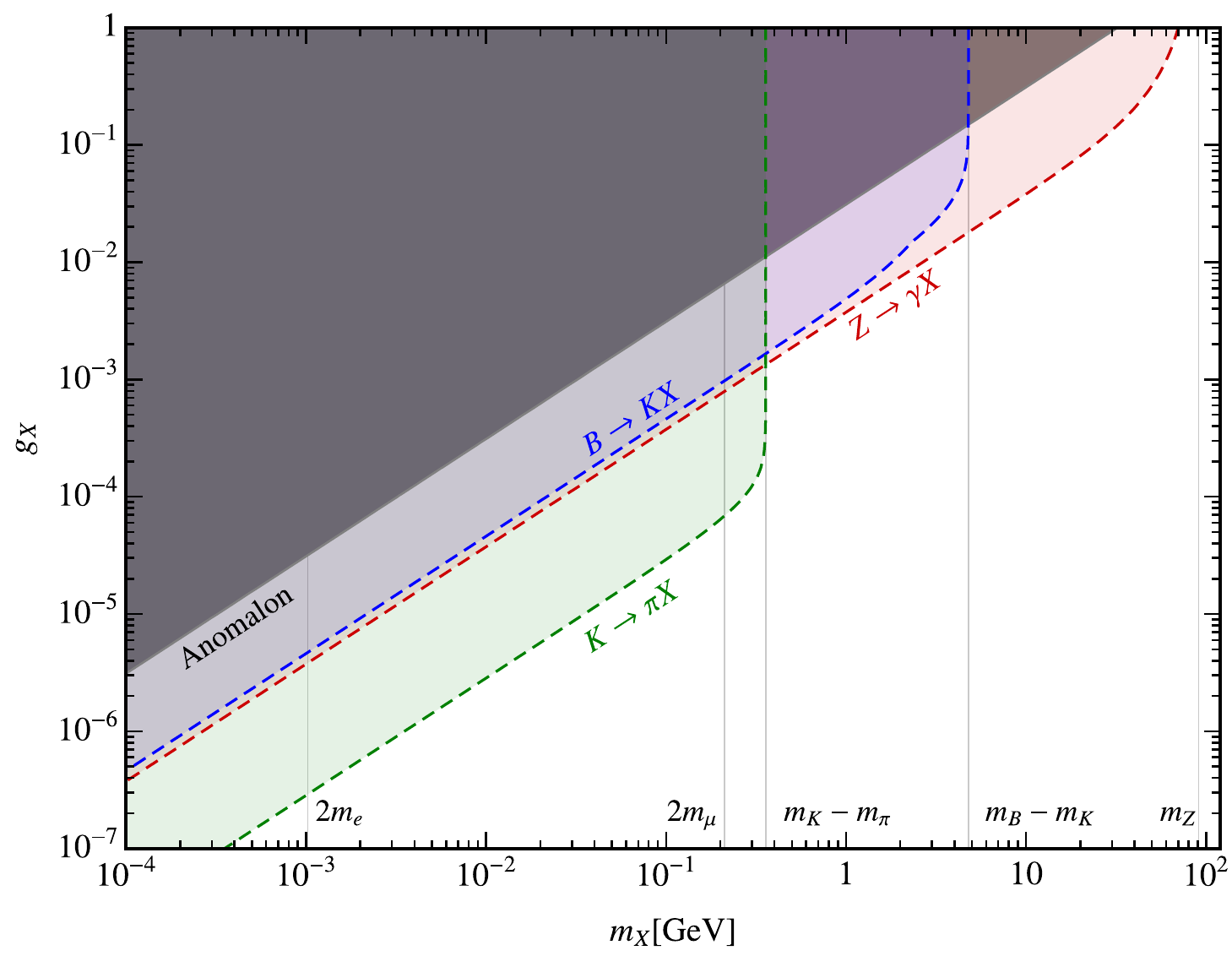}
		\caption{
			{\em Left panel:}
		Constraints on a vector $X$ coupling to baryon number, assuming
		a kinetic mixing with the SM photon $\epsilon \sim e g_X / (4 \pi)^2$, and no additional invisible $X$ decay channels.
		Colored regions with solid borders indicate constraints from visible
		decays, dashed borders correspond to missing
		energy searches, dot-dashed borders denote displaced limits, and dotted borders denote projections based on current expected sensitivities.
		The gray regions indicate constraints from the previous literature.
		The new constraints come from searches for
		$ K \rightarrow \pi X $
(green)~\cite{AlaviHarati:2003mr,Anisimovsky:2004hr,Artamonov:2008qb}, $ B
\rightarrow K X $ (blue)\cite{Aaij:2012vr,Aubert:2008ps,Olive:2016xmw,Grygier:2017tzo}, $ Z \rightarrow X \gamma $
(red)\cite{Acciarri:1997im,Abreu:1996pa,Adriani:1992zm,Acton:1991dq,Adeva:1991dw}, and very displaced decays at the CHARM proton
beam dump experiment~\cite{Bergsma:1985qz}.
		For the latter, the enhanced $K \rightarrow \pi X$ decays
		result in larger $X$ production than computed in naive
		analyses~\cite{Alekhin:2015byh,Gardner:2015wea}.
		The `anomalon' line shows the approximate region in which
		anomaly-cancelling fermions would be light enough to have been
		detected~\cite{Dobrescu:2014fca}.
		The other gray constraints are (left to right) from
$ \phi $ and $ \eta $ decays~\cite{Tulin:2014tya}, and $\Upsilon$
		decays~\cite{Carone:1994aa}. Improved `tree-level' limits
		are expected from photoproduction of $X$ at the GlueX
		experiment~\cite{Fanelli:2016utb}.
		{\em Right panel:} As above, but with the assumption that $X$ dominantly decays invisibly.
		}
		\label{fig:bvec2}
\end{figure*}

Figure~\ref{fig:bvec2} shows a selection of experimental bounds on
the coupling of a baryon number vector, including the anomalous
processes described above and non-enhanced processes.~\footnote{For the
limits throughout we assume that decays beyond $ 10 $~m ($ 4 $~m) are
considered invisible for the high energy (flavor) limits. Furthermore,
we assume decays within 1~mm (5~mm) are counted as prompt for high
energy (flavor) measurements.} Here we are assuming a UV completion
that preserves the EW symmetry, in order to evade electroweak precision
and direct collider constraints (see section~\ref{sec:UV}). As this
figure indicates, including the anomalous processes gives a significant
improvement in the constraints across a wide mass range.
To reiterate, these anomalous-coupling-based bounds arise whenever the
mixed electroweak-$ U(1)_X $ anomalies are cancelled by heavy fermions
whose masses do not (dominantly) arise from EW-symmetry breaking. 
The limits displayed in Figure~\ref{fig:bvec2}
correspond to the UV completion introducing no extra EW-symmetry
breaking; if the heavy fermions receive some fraction of their
mass from an EW-symmetry breaking contribution, then the limits
will be reduced proportionately. Significantly weakening these bounds
would require either more exotic UV
completions which do not introduce new fermions, or
for the new fermions to have dominantly EW-breaking masses, with the latter option running into strong
observational constraints (see e.g.\ Appendix~\ref{ap:chiral}). 



For the couplings shown in Figure~\ref{fig:bvec2},
the decay time of $X$ is $\ll 1 \sec$ for $m_X > 2 m_e$, so for $m_X \gtrsim 10 \MeV$, 
$X$ will have a strongly Boltzmann-suppressed abundance
in the early universe at temperatures $\lesssim 2 \MeV$, when 
non-equilibrium processes are important.
For $m_X < 2 m_e$, the lifetime will be long enough for
there to be constraints from the late-time decay of an
$X$ abundance produced in the hot early universe,
while for $2 m_e < m_X \lesssim 10 \MeV$, there will be constraints,
at large enough $g_X$, from $X$ freeze-out occurring after SM
neutrino freeze-out~\cite{Nollett:2013pwa}.
We leave the calculation of such constraints to future work.
Supernova cooling bounds~\cite{Raffelt:1996wa}
will also constrain some range of parameters; however, the emission rates
computed in the literature~\cite{Grifols:1988fv,Rrapaj:2015wgs} do not
take into account either in-medium mixing effects~\cite{Hardy:2016kme}
or anomalous production, so would need to be re-calculated.

The right-hand panel of Figure~\ref{fig:bvec2} shows
 the constraints that
arise if $X$ has a significant branching ratio to invisible states.
For example, one light Dirac fermion $\chi$ with $X$-charge of 1 and $2m_\chi      
<m_X$ will result in an invisible branching fraction of $\gtrsim 30 \% $.
The constraints from missing energy searches are strong,
and limit the discovery prospects for light dark matter
coupled through such a mediator at neutrino experiments~\cite{Aguilar-Arevalo:2017mqx,Frugiuele:2017zvx}.
A small modification of this plot will also apply
to a vector coupled to lepton number. This has the same anomalous couplings
as a baryon number vector, but its branching ratio to neutrinos
is larger than 30\% everywhere (and is $\sim 100 \%$ for $m_X < 2 m_e$).
In addition to the anomalous bounds, a lepton number coupled vector
will also be constrained by neutrino-electron scattering experiments
(section~\ref{sec:noncol}), which provide the dominant constraints
for $m_X \gtrsim m_K - m_\pi$.

\subsubsection{Comparison to literature}

Some early papers on baryon number vectors, such as~\cite{Carone:1994aa,Carone:1995pu},
considered models in which the anomalies are cancelled by new SM-chiral 
fermions, and noted that these could be made compatible with the electroweak precision (EWP) constraints
of the time.
At least one later paper~\cite{Dobroliubov:1991qz} considered
$X$ production through its anomalous couplings to SM gauge bosons
(presumably assuming a model where anomalies are cancelled at a higher scale),
but performed the calculation incorrectly.\footnote{In particular, 
\cite{Dobroliubov:1991qz} states that \emph{``The
  contribution of a fermion does not depend on its mass while the latter
  is much smaller than $m_Z$, and its contribution is suppressed with
  factor $\simeq 0.1 (m_Z/m_f)^2$ for $m_f \gtrsim m_Z$ (so we do not
  take the top quark into account).''}
  As per section~\ref{sec:xlamps}, since the $X$ couples vectorially to SM fermions,
  the amplitude for longitudinal $X$ emission does not depend on the mass
  of the SM fermions in the triangle diagrams. 
  Ref.\ \cite{Dobroliubov:1991qz}'s calculations would, for example, imply an anomalous
  amplitude for longitudinal $B-L$ vector emission,
  which cannot be present.}

More recent papers, such as~\cite{Tulin:2014tya,Feng:2016ysn}, consider models in
which the anomalies are cancelled at a higher scale in a SM-preserving
way, but ignore the anomalous couplings in the low-energy theory.
\cite{Dobrescu:2014fca} considers such a model, and studies the direct
production of the heavy fermions at colliders. The constraints derived
are, at $m_X \lesssim m_K$, significantly weaker than
those from the anomalous couplings present in such models.
This is in analogy to searches for axion-like particles,
in which it is often easier to detect effects involving the light degree
of freedom than it is to probe the heavy states beyond the cutoff.


A number of works have looked at the consequences of
`effective Chern-Simons' interactions of a new vector with
the SM, i.e.\ couplings of the form $\epsilon^{\mu\nu\rho\sigma} X_\mu
Z_\nu \partial_\lambda A_\rho$ etc to the SM EW gauge bosons, as
reviewed in \cite{Alekhin:2015byh}. Some of these works, such
as \cite{Antoniadis:2009ze} and \cite{Alekhin:2015byh}, seem to claim
that unsuppressed SM-violating values for these couplings can occur in the SM + $X$ EFT,
due to heavy anomaly-cancelling fermions which dominantly obtain
masses from a SM-singlet VEV. As we have discussed, 
 the values of such
WZ terms are determined,
up to $m_{\rm{EW}}^2/m_f^2$ corrections,
by the couplings of $X$ to SM fermions,
within that class of UV completions.
Most works other than~\cite{Alekhin:2015byh} also consider $m_X > m_Z$,
giving rather different phenomenology.
For low $m_X$, the FCNC constraints we have derived dominate
those given in~\cite{Alekhin:2015byh}.

An interesting side point is that the kind of anomalous interactions
we have considered for a new light vector may also have analogues
within the SM. \cite{Harvey:2007rd,Harvey:2007ca} propose that the presence
of the $\omega$
meson in the low-energy theory of QCD, which behaves analogously to a baryon number coupled vector,
should require a WZ term of the form $\epsilon^{\mu\nu\rho\sigma} \omega_\mu
Z_\nu F_{\rho\sigma}$, which in the low-energy theory
would result in a $\omega \gamma \bar\nu \nu$ coupling.
Since the $\omega$ meson is composite at scales comparable to its
mass, high-energy $E^2/m_\omega^2$ enhanced processes of the type we have
considered will not apply.


\section{Tree-level breaking}
\label{sec:tree}

In addition to being broken through chiral anomalies,
as considered in the previous section, the SM current that a light
vector is coupled to may also be broken at tree level,
within the SM + $X$ EFT. The possible sources of breaking
via dimension-4 couplings are:

\begin{itemize}

	\item \emph{Axial couplings to SM fermions}: since
fermion Dirac masses break
axial symmetry, an axially-coupled vector $X$ will have
longitudinal $X$ production processes enhanced by $(m_f/m_X)^2$,
		where $m_f$ is the mass of the relevant SM fermion.\footnote{
			The QCD chiral condensate also breaks chiral symmetry,
			so axial coupling processes involving hadrons
			such as nucleons are not suppressed by the small
			first-generation quark masses, but only by the nucleon mass etc.}
		Axial couplings
also allow $X$ to have `anomalous' two-photon and two-gluon couplings
--- in fact, as we will see in section~\ref{sec:gluefuse}, they render
such couplings unavoidable, due to the dependence of triangle
amplitudes on SM fermion masses.
Consequently, processes involving heavy SM fermions, or anomalous
		couplings, can provide strong constraints on axial couplings.\footnote{
If SM neutrinos are Majorana, then there will also be $U(1)_X$ breaking
effects from couplings to neutrinos, suppressed by the small neutrino masses; we will ignore 
		these.}

\item \emph{Generation-non-universal couplings}:
		due to the non-diagonal quark mass matrices,
		generation-non-universal couplings of a vector
		to quarks generically lead to tree-level FCNCs
		(see Appendix~\ref{ap:treefcnc}), which are
		tightly constrained by experiment.
		If the vector is light, then these vertices can lead
		to flavor-changing quark decays via the emission
		of a real longitudinal $X$, placing even stronger constraints
		on such operators.
		Even without tree-level FCNCs, generation-non-universal
		couplings in combination with SM flavor-changing vertices
		generally result in $X$-non-conserving processes (section~\ref{sec:penguin}).
		Lepton flavor-changing vertices are also subject to strong
		experimental constraints,
		but can be more easily avoided. 
		Given the strong experimental constraints on tree-level flavor-changing
		couplings, we will assume throughout this paper that these are
		highly suppressed.

	\item \emph{Weak-isospin violation}: $W \bar{q}_u q_d$ or $W l \bar{\nu} $ vertices
		may be $U(1)_X$-breaking, if $X$ couples differently to left-handed fermions
		that are members of the same $SU(2)_L$ doublet, and does
		not have the compensating coupling to the $W$. $X_L$ radiation from charged current processes is then enhanced. 

	\item \emph{EW couplings}: in addition to its couplings to
		fermions, $X$ may have dimension-4 couplings to
		the SM EW gauge bosons, or to the SM Higgs.
		A simple example is mass mixing with the $Z$ boson~\cite{Davoudiasl:2012ag,Kahn:2016vjr}, which
		leads to a $XWW$ coupling. For simplicity, we
		will only consider Wess-Zumino type couplings in this paper; 
		as discussed above, these are determined by the SM
		fermion couplings of $X$, if the UV completion does not introduce
		extra EWSB, and are suppressed by a loop factor.

              \end{itemize}

In this section, we will identify various process which place
strong constraints on couplings of these forms.
To keep the discussion as general as possible, we parameterize the interaction between $X$ and SM fermions as
\begin{equation}
{\cal L}  \supset g_X X_\mu \sum_i \bar\psi_i \gamma^\mu (c_i^V + c_i^A \gamma_5) \psi_i\,.
\end{equation}

\subsection{Radiation from axial current}
\label{sec:axialrad}
The simplest such process is the radiation 
of a longitudinal $X$ from a massive SM fermion via its axial coupling.
This is enhanced by $(m_f / m_X)^2$, so it is 
advantageous to use heavy quarks or leptons
in these searches. 

As a relevant example, consider heavy quarkonium decays such as $ \Upsilon \rightarrow \gamma X $. Here, the coupling of $X$ to the $b$ quark leads 
to enhanced $X_L$ emission. 
This process has been considered in~\cite{Fayet:2006sp,Fayet:2007ua} (and in \cite{Dermisek:2006py}
for the case of a light pseudoscalar), and has a branching ratio of
\begin{equation}
	{\rm Br}(\Upsilon_{1S} \rightarrow \gamma X) \simeq 4 \times 10^{-5} |c^A_b|^2
	\left(\frac{\TeV}{f_X}\right)^2\,.
\end{equation}
The $\Upsilon \rightarrow \gamma$ + invisible~\cite{delAmoSanchez:2010ac}
and $\Upsilon \rightarrow \gamma (X \rightarrow \mu^+ \mu^-)$~\cite{Love:2008aa}
decays
are both measured to have small ($< 10^{-5}$) branching ratios,
giving constraints on $g_X c^A_b$. Similar constraints can be formulated for the axial charm coupling through $ J / \psi $ decays (using, e.g., constraints from~\cite{Ablikim:2015voa}).

An axial coupling to the top quark gives the largest $m_f^2/m_X^2$ enhancement.
However, the much smaller number of top quarks produced in experiments
means that, at small $m_X$, axial couplings to lighter quarks will
give stronger constraints, while at larger $m_X$, other high-energy production
mechanisms will generally dominate.

As we discuss in the next section,
the constraints
on quark axial couplings from FCNC meson decays
generally dominate those from other processes,
such as $\Upsilon$ decays, except in the case of
purely right-handed down-type couplings.


\subsection{FCNCs}
\label{sec:penguin}

As reviewed in section~\ref{sec:anomfcnc}, flavor-changing transitions
between down-type quarks can proceed via a $W$-boson / up-type-quark
loop.
If $X$ couples to a quark current that is broken at tree level,
then the effective $d_i d_j X$ vertex obtains
a value $\sim \frac{1}{16\pi^2} g^2 g_X \times $ (CKM elements).
Compared to the currents broken by the chiral anomaly considered in
section~\ref{sec:anomfcnc}, the lack of additional loop suppression
means that even stronger constraints can be obtained from
flavor-changing meson decays.
This is in precise analogy to meson decays via an axion-like particle
with couplings to quarks~\cite{Dolan:2014ska}, compared to one with a $W \tilde{W}$
coupling~\cite{Izaguirre:2016dfi}.

Within the SM + $X$ EFT, flavor-changing penguin diagrams
involving the $X$ coupling to quarks are divergent.
Consequently, unless the sum of these divergences
cancels, there must be flavor-changing $d_i d_j X$
counterterms in the EFT. 
Thus, unlike in section~\ref{sec:anomfcnc},
we cannot assume that the UV theory contributes no unsuppressed $d_i d_j X$ FCNCs.
However, we can estimate the contributions from `simple' UV completions
by evaluating the logarithmically divergent $d_i d_j X$ amplitudes
within the EFT, and assuming that these are resolved by UV states,
giving an overall $\log(M / m_{\rm EW})$ amplitude (where $M$ is the UV scale).

This prescription is complicated by the fact that the divergent
parts of EFT amplitudes are not gauge-independent; however, their $m_t^2/m_W^2$-enhanced parts are.\footnote{This is only within the family of $R_\xi$ gauges; the situation is more complicated
in unitary gauge. In fact even within the SM, off-shell $d_i d_j Z$ amplitudes
are divergent in unitary gauge, with divergences only cancelling
when combined with $W^+ W^-$ box diagrams, or when the $Z$ is put
on-shell~\cite{He:2009rz}.
	}
	These give an effective $d_i d_j X$ coupling of
\begin{align}
	\label{eq:eftax}
	&g_{X d_i d_j}^A \simeq
	\frac{1}{16 \pi^2} g^2 g_X C_{t d_i d_j} V_{td_i} V^*_{td_j} + \dots \,,\\
	&C_{t d_i d_j} = \frac{1}{2} \frac{m_t^2}{m_W^2} 
	(c^L_{d_i} + c^L_{d_j} - 2 c^R_t) \log \frac{M^2}{m_t^2} + \dots \notag
\end{align}
(where we have neglected the down-type quark masses).
In the case of universal vectorial couplings, this vanishes, as expected
from a coupling to a conserved current.
Below, we will see how an effective coupling of this kind arises in
a two Higgs doublet model (2HDM) UV completion, where the UV states are the heavy charged Higgses.
It is of course possible that other UV completions may lead to further
cancellations; unlike for the anomalous $XWW$ coupling,
evaluating FCNC amplitudes properly requires knowing the full theory.

Interestingly, the FCNC vertex can be enhanced by the top quark mass (as opposed to suppressed by a light quark mass) even in the case where $ X $ does not couple to the top quark at all.
For example, left-handed couplings to down-type quarks can also give a $m_t^2/m_W^2$ enhanced vertex, due to the self-energy diagrams. 

As discussed in Appendix~\ref{ap:treefcnc}, generation-non-universal
couplings to quarks can lead to tree-level flavor-changing vertices.
In particular, if $X$ couples to both up- and down-type left-handed quarks, then
either these couplings are related in a specific weak-isospin-violating
way, or there are tree-level FCNCs. 
From equation~\ref{eq:eftax}, we see that unless the left-handed down-type
couplings are equal, then there will be some down-type FCNCs 
which are unsuppressed by small quark masses.
Thus, the safest way to have generation-non-universal couplings
to quarks, without introducing dangerous FCNCs, is generally
to have dominantly right-handed couplings.

Note that we assumed above that $ X $ does not mix with the $ Z $. If there is such a mixing, the resulting $XWW$ coupling gives an
additional contribution to FCNCs --- in Feynman gauge, the coupling
of $X$ to the charged Goldstones will give a $m_t^2/m_W^2$ enhanced, logarithmically-divergent, contribution~\cite{Inami:1980fz}.

As discussed in section~\ref{sec:fcncexp}, experimental searches
for FCNC meson decays via these effective operators will depend on how
$X$ decays. Compared to the effective operators arising from 
anomalous $XWW$ couplings, the lack of an extra loop suppression factor
in equation~\ref{eq:eftax} results in larger branching ratios for a given
coupling, so gives stronger constraints on $g_X$.


\subsubsection{UV completion}
\label{sec:penguinuv}

\begin{figure}
          \begin{center} 
\includegraphics[width=\columnwidth]{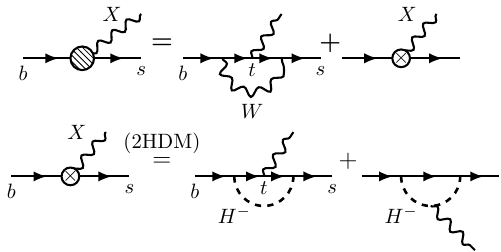} 
\end{center}
\caption{
			\emph{Top row}: effective $bsX$ FCNC vertex for a vector with
			right-handed couplings to quarks, obtained by integrating 
			out the $W$.
			The $W$ loop diagram is divergent, so must be cancelled by a tree-level
			counter-term in the EFT.
			In a UV completion, this counterterm will arise from
			e.g.\ integrating out loops involving heavy states.
			Couplings to external quark legs (`self-energy' diagrams)
			are omitted, since these are suppressed by down-type quark
			masses for right-handed couplings.
			\emph{Bottom row}: calculation of the EFT
			FCNC counterterm in the 2HDM UV completion discussed in
			section~\ref{sec:penguinuv}, where it corresponds to diagrams
			with charged Higgs loops.
			Within the UV theory, these cancel the log divergence
			of the $W$ (Goldstone boson) loop diagrams, and give a
			$\log(m_{H^\pm}^2/m_t^2)$-enhanced amplitude.
			See section~\ref{sec:penguinuv} for details.
		}
		\label{fig:penguin}
\end{figure}

It can be enlightening to see how the effective vertex
arises within a particular UV completion. Here, 
we calculate the $d_i d_j X$ vertex within the 2HDM $U(1)_R$ UV completion
presented in~\cite{Kahn:2016vjr}. In their UV theory,
$X$ has purely right-handed couplings to SM fermions,
while there are Yukawa terms,
\begin{equation}
	\LL \supset Y_u H_u Q u_R + Y_d H_d Q d_R + Y_l H_d L l_R + {\rm h.c.}\,,
\end{equation}
along with new anomaly-cancelling fermions which get masses from
a pair of SM-singlet, but $X$-breaking, VEVs. Since $H_u$ and $H_d$ have
$X$ charges, the SM fermions can gain comparable vector and axial couplings to $X$.
As illustrated in Figure~\ref{fig:penguin}, calculating the $d_i d_j X$ vertex
in this model involves adding the charged Higgs exchange diagrams,
which cancel the logarithmic divergences in the $W$ exchange diagrams.
Keeping only the $\frac{m_t^2}{m_W^2}\log \frac{m_{H^\pm}^2}{m_t^2}$
enhanced terms,
and choosing the 2HDM parameters so that $X$ does not mix with $Z$
(i.e.\ $q_{H_u} s_\beta^2 = q_{H_d} c_\beta^2$, in the notation
of~\cite{Kahn:2016vjr}),
we obtain the effective vertex,
\begin{equation}
	g^{{\rm 2HDM}}_{X d_i d_j}
	= \frac{1}{16 \pi^2} g_X V_{tb} V_{ts}^* (q_{H_u} + q_{H_d}) c_\beta^2 s_\beta^2
	\frac{1}{2} Y_t^2 \log \frac{m_{H^\pm}^2}{m_t^2}\,.
\end{equation}
Since the $u_R$ coupling of $X$ in this model is $C^R_u
= -\frac{1}{2} g_X q_{H_u} = - \frac{1}{2} g_X (q_{H_u} + q_{H_d})
c_\beta^2$, and $m_t = Y_t s_\beta v / \sqrt{2}$, this gives,
\begin{equation}
	g^{{\rm 2HDM}}_{X d_i d_j}
	= - \frac{1}{16 \pi^2} c_R^u V_{tb} V_{ts}^* g^2 g_X
	\frac{m_t^2}{m_W^2} \log \frac{m_{H^\pm}^2}{m_t^2}\,.
\end{equation}
As expected, this matches the form of the log-divergent terms
in the EFT calculation given by equation \eqref{eq:eftax}, with the UV scale set by the mass of the charged
Higgs states.
In a realistic 2HDM model, this scale may not be so far above 
the top mass (see e.g.~\cite{Eberhardt:2013uba}), and non-log-enhanced terms may be numerically important. However, in the absence of fine-tuned cancellations,
the log-enhanced term should be a good parametric estimate of the
effective vertex, and illustrates how the EFT result corresponds to a
UV computation. 

It is possible that other UV completions could lead to cancellations
that suppress FCNCs below the level expected from equation~\ref{eq:eftax}.
However, we are not aware of models where this occurs generically,
without additional fine-tuning.

\subsection{Triangle diagram amplitudes}
\label{sec:gluefuse}

A new vector with purely vectorial couplings to SM fermions can
only have have anomalous couplings of the $\propto \varphi (g^2 W^a
\tilde W^a - g'^2 B \tilde B)$ form, since the photon and the gluon both
have purely vectorial couplings as well. However, an axially-coupled vector can
have anomalous couplings to $F \tilde{F}$ and $G \tilde{G}$.
In particular, as reviewed in Appendix~\ref{ap:anom}, 
the $X_L$ coupling from fermion triangle diagrams has a fermion-mass-dependent
piece proportional to the axial coupling of $X$ to the fermion.
Consequently, even if the chiral fermion content of the SM + $X$ EFT
is non-anomalous, the differing masses of the SM fermions
will give enhanced $X_L$ production, unless the external momenta
are small or large compared to all of the relevant fermions masses.

While the $G \tilde{G}$ coupling does result in enhanced
production of high-$p_T$ $X$ at the LHC (resulting in monojet
bounds, if $X$ registers as missing energy), the increase
of the gluon parton distribution function at low energies more than compensates for
the falling $({\rm energy} / m_X)^2$ enhancement. Consequently,
$X$ production from gluons is actually dominated by gluons
of the minimum allowed energies; if $m_X \gtrsim \Lambda_{\rm QCD}$,
there is no parametric longitudinal mode enhancement
for the overall production rate.

As reviewed in section~\ref{sec:noncol}, the $F \tilde{F}$ coupling
leads, for light enough $X$, to Primakoff process production
in stars~\cite{Raffelt:2006cw}, giving constraints from stellar energy loss arguments.


\subsection{Radiation from charged current decays}
\label{sec:ccrad}
In models with weak-isospin breaking, the $W \bar{q}_u q_d$ or $Wl  \bar{ \nu}$ vertices may violate $U(1)_X$. This leads to charged current decays that radiate $X_L$ at an
$({\rm energy} / m_X)^2$ enhanced rate. The simplest example is
 $W \rightarrow l \nu X$ decays~\cite{Karshenboim:2014tka}.
While using branching ratios places weak constraints
on the coupling of $X$ ($f_X \gtrsim 10 \GeV$),
it is possible that `bump-hunt' searches in the invariant mass
of the $X$ decay products could give better sensitivity.

For $W l \bar{\nu}$ couplings, the 
$\pi^+ \rightarrow l^+ \nu_l X$ decay, and the
analogous charged kaon decay, are potentially
sensitive channels.
If $X$ decays to $e^+ e^-$, then the leading competing SM decay
is $\pi^+ \rightarrow e^+ \nu_e (\gamma^* \rightarrow e^+ e^-)$.
At tree level in chiral perturbation theory, the SM decay is
helicity suppressed, since the current that the pion couples to is
conserved in the massless-lepton limit. In contrast, since $X$ couples
differently to the electron and the neutrino, the helicity suppression
is lifted in the $\pi^+ \rightarrow e^+ \nu_e X$ decay, which is
therefore enhanced by $\sim m_\pi^4 / (m_e^2 f_X^2)$ compared to the SM
decay.

Experimentally, searches for the $\pi^+ \rightarrow
e^+ \nu_e (h \rightarrow e^+e^-)$ decay, where $h$ is a light scalar Higgs,
constrain the branching ratio of this decay to be
$\lesssim 10^{-9}$~\cite{Egli:1989vu} for $m_h \gtrsim 10 \MeV$ (for comparison,
the $\pi^+ \rightarrow e^+ \nu_e e^+ e^-$ branching ratio
is measured to be $(3.2 \pm 0.5) \times 10^{-9}$~\cite{Egli:1989vu}).
Taking a toy model where $X$ couples only to electrons, and not
to pions or neutrinos (e.g.\ as occurs in the (modified) `$B-L$' model
of~\cite{Feng:2016ysn}, discussed in section~\ref{sec:phenomodels}),
the $\pi^+ \rightarrow e^+\nu_e X$
branching ratio is
\begin{equation}
	{\rm Br}_{\pi^+ \rightarrow e^+\nu_e X} \simeq 10^{-9}
	\left(\frac{200 \GeV}{f_X}\right)^2\,.
\end{equation}


\subsection{Non-collider constraints}
\label{sec:noncol}

At smaller masses and couplings,
stellar energy loss arguments~\cite{Raffelt:2006cw},
and bounds on the decay of an $X$ abundance produced
in the hot early universe~\cite{Cadamuro:2011fd}, will
provide additional constraints (c.f.\ the discussion
in section~\ref{sec:bvector})
If longitudinal mode production is comparable
to or dominates transverse production, as will be
true in many cases, these constraints
will be analogous to those for ALPs.
For $m_X < 2 m_e$, the allowed
SM decays of $X$ are to neutrinos, or the (very slow) loop-induced
decay to more than two photons --- consequently, the bounds from late-time cosmological
decays will be significantly different to the ALP case,
where the two-photon decay is allowed.
Additionally, if the vectorial couplings of $X$ are significantly larger
than the axial ones, then the $X$ decay rate will be mainly set by
the vectorial couplings, changing the relation between production
and decay rates.

Stellar energy loss  bounds give~\cite{Raffelt:2006rj}
\begin{align}
	\nonumber	c^A_e f_X^{-1} &\lesssim (10^9 \GeV)^{-1} \,,\\
	c^A_N f_X^{-1} &\lesssim (10^8 \GeV)^{-1} \,,
	\label{cool}\\
	\nonumber \Am_{X\gamma\gamma} f_X^{-1} &\lesssim (5 \times 10^7 \GeV)^{-1}\,,
\end{align}
where $c^A_N$ is the axial coupling to nucleons
--- these are simply translations of the coupling bounds for an axion-like particle,
applied to longitudinal $X$ emission.
Bounds from horizontal branch and red giant stars,
which constrain $c^A_e$ and $\Am_{X\gamma\gamma}$, apply
at $m_X \lesssim 10 \MeV$, while those from SN1987A,
which constrain $c^A_N$ and $\Am_{X\gamma\gamma}$,
apply at $m_X \lesssim 100 \MeV$.
Supernova cooling bounds and cosmological
decay constraints do not apply at large enough couplings,
when the vector interacts too strongly / decays too fast
(though as per section~\ref{sec:bvector}, there will still be cosmological
constraints at small enough vector masses).

These astrophysical bounds have strong implications for searches
for a fifth force mediated by an axially-coupled $X$.
In particular, the spin-spin interactions induced by the exchange
of $X$~\cite{Dobrescu:2006au,Karshenboim:2010cg,Hunter:2013hza}
are far better constrained indirectly through (\ref{cool}),
than they are by laboratory searches for spin-dependent
forces.


In addition to constraints coming from the ALP-like behaviour
of longitudinal modes, there are also constraints which are
based simply on the form of the vector's coupling to fermions.
For example, if the products $c^A_e c^V_q$ or $c^V_e c^A_q$ are non-zero,
then $X$ exchange leads to atomic parity violation (APV),
with the $c^A_e c^V_q$ term being much more strongly constrained
due to the coherent coupling to the nucleus. 
Measurements of the effective `weak charge' of $^{133}{\rm Cs}$ match the SM value to 
better than the percent level~\cite{Wood:1997zq,Porsev:2009pr}.
For $m_X \gtrsim 3 \MeV$, the $X$ vector effectively mediates
a contact-operator interaction between the nucleus and the electrons,
giving a constraint of the form,~\cite{Bouchiat:2004sp}
\begin{equation}
	\left| 0.47 c_u^V c_e^A+ 0.53 c_d^V c_e^A\right|  ^{1/2} f_X^{-1} \lesssim (10 \TeV)^{-1}\,.
  \end{equation} 

Also, if $X$ couples to neutrinos and electrons, then
$\nu - e$ scattering measurements place strong constraints
on the product of the $\nu$ and $e$ couplings.
Again, for $m_X$ greater than the scattering momentum transfer,
which is generally between $1 - 100 \MeV$, these 
give constraints of the form $c_e c_\nu f_X^{-1} \lesssim \TeV^{-1}$
\cite{Bilmis:2015lja,Jeong:2015bbi}.




\subsection{Comparison of constraints}

To compare our new constraints to each other, and to existing bounds,
we need to choose a specific model for the couplings of the new vector. A
convenient choice is the $U(1)_R$ 2HDM model of~\cite{Kahn:2016vjr}, as
discussed in section~\ref{sec:penguinuv}. This provides a UV-complete
model, in which the only light new state is a vector with chiral couplings to the SM
fermions. 
The SM + $X$ EFT within this model is anomalous,
with the anomaly cancelled in the UV by heavy fermions
getting their mass from a SM-singlet VEV.
For simplicity, we choose the model parameters such that there
is no $Z$-$X$ mass mixing --- this means that the couplings to SM fermions are purely right-handed,
apart from loop-induced kinetic mixing contributions. 

\begin{figure*}
	\begin{center}
		\includegraphics[width=0.47\textwidth]{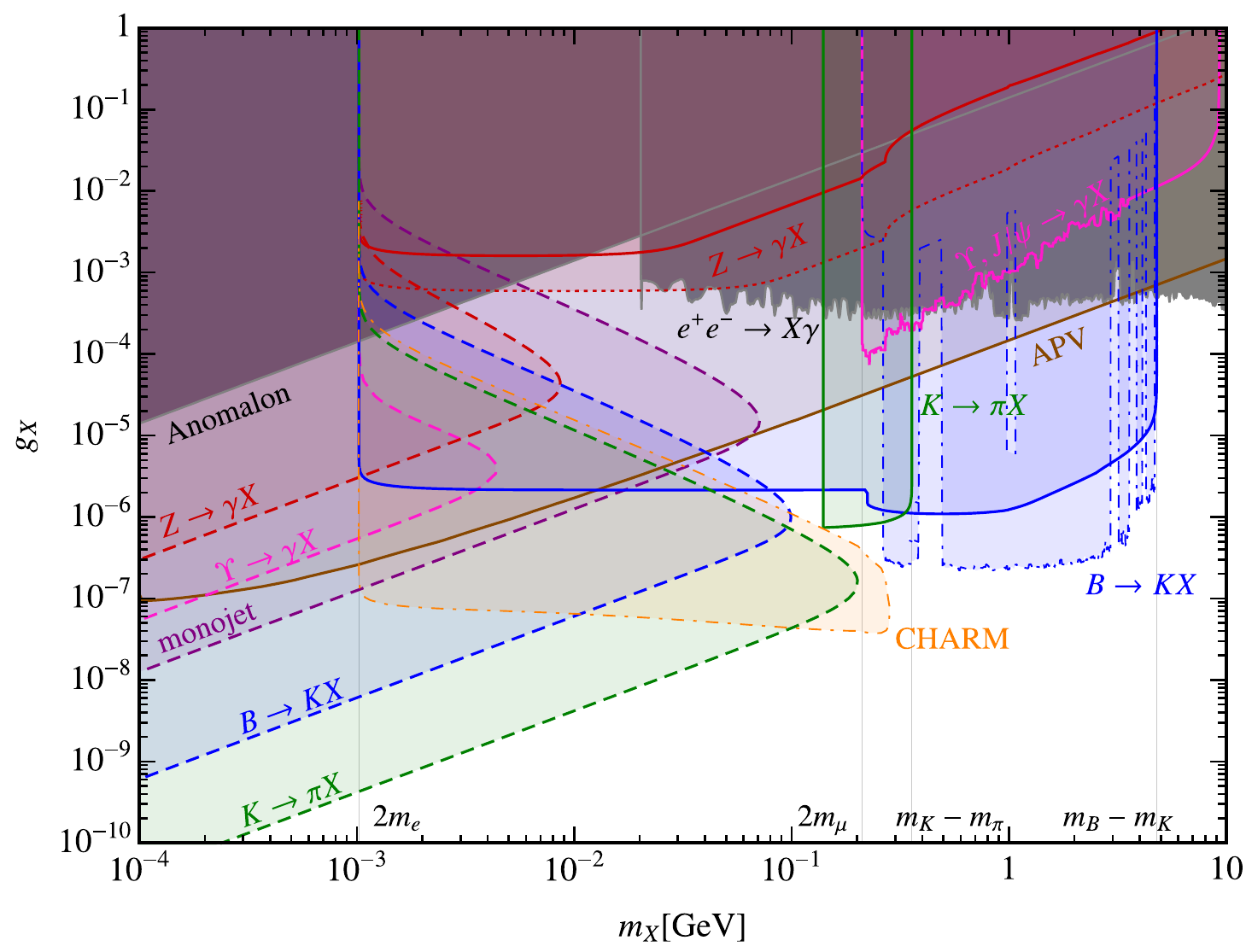}
		\includegraphics[width=0.47\textwidth]{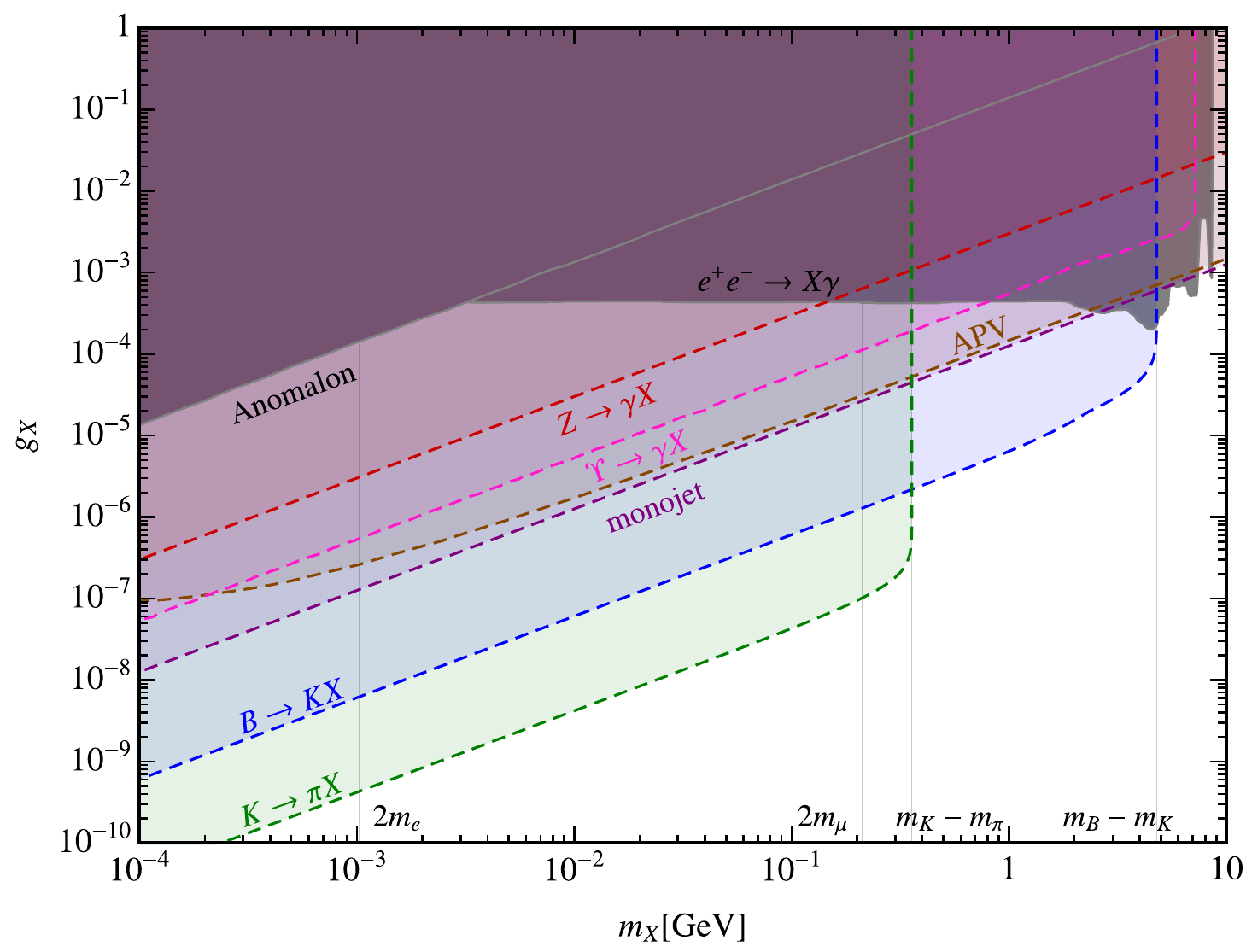}
		\caption{
			{\em Left panel:}
			Constraints on a vector $X$ with generation-universal couplings to right-handed
			quarks and leptons, assuming no additional invisible $X$ decay channels.
		Colored regions with solid borders indicate constraints from visible
		decays, dashed borders correspond to missing
		energy searches, the dot-dashed borders indicate displaced searches, and dotted borders denote projections based on current expected sensitivities.
		The gray regions indicate constraints which do not utilitize the energy-enhanced production of the longitudinal mode.
		The new constraints come from searches for
		$ K \rightarrow \pi X $
(green)~\cite{AlaviHarati:2003mr,Anisimovsky:2004hr,Artamonov:2008qb}, $ B
\rightarrow K X $ (blue)\cite{Aaij:2012vr,Aubert:2008ps,Olive:2016xmw,Grygier:2017tzo,Aaij:2016qsm}, $ Z \rightarrow X \gamma $
(red)\cite{Acciarri:1997im,Abreu:1996pa,Adriani:1992zm,Acton:1991dq,Adeva:1991dw}, very displaced decays at the CHARM proton
beam dump experiment~\cite{Bergsma:1985qz}, and monojets~\cite{CMS-PAS-EXO-16-048}.
		The enhanced $K \rightarrow \pi X$ decays
		result in larger $X$ production than computed in naive
		analyses~\cite{Alekhin:2015byh,Gardner:2015wea}. The $ \Upsilon $ and $ J /\psi $ constraints use~\cite{delAmoSanchez:2010ac,Love:2008aa,Ablikim:2015voa}.
		The `anomalon' line shows the approximate region in which
		anomaly-cancelling fermions would be light enough to have been
		detected~\cite{Dobrescu:2014fca} and the $ e ^+ e ^- \rightarrow X \gamma $ constraint is from a search for dark photons~\cite{Lees:2014xha}.
		{\em Right panel:} As above, but with the assumption that $X$ dominantly decays invisibly. 
		}
		\label{fig:axial1}
	\end{center}
\end{figure*}

Figure~\ref{fig:axial1} illustrates a selection of the constraints
we have discussed above, applied to the $U(1)_R$ model. Compared
to the APV constraints, which are the most stringent in the existing literature for our parameters, the
new constraints are significantly stronger across a wide mass range.
In particular, the $X$ production rate at colliders and in proton beam
dump experiments is
significantly larger than the predictions of e.g.~\cite{Kahn:2016vjr},
which did not take into account $X_L$ production in meson decays.
Compared to the analyses of longitudinal mode production by
Fayet~\cite{Fayet:2006sp,Fayet:2007ua}, we obtain improved
constraints by calculating penguin diagram contributions to
flavor-changing meson decays, which are parametrically
larger than other new contributions.

For $m_X > 2 m_e$, the decay times for the couplings shown in the plot
are $\ll 1 \sec$, so the discussion of cosmological
bounds in section~\ref{sec:bvector}
applies. Supernova cooling bounds will constraints some region of couplings,
though we leave computing these to future work.

The form of Figure~\ref{fig:axial1} depends on the similar quark
and lepton couplings in the $U(1)_R$ model. The constraints
on a vector with leptophilic or leptophobic couplings will
fit together differently, and some of the processes
that we do not show in Figure~\ref{fig:axial1}, such
as 
$\pi^+ \rightarrow e^+ \nu X$,
can become important.
Furthermore, if $X$ has dominantly invisible decays, the various missing energy searches
will extend up to higher $m_X$, again resulting in strong coupling constraints,
as illustrated in the right-hand panel of Figure~\ref{fig:axial1}.

\section{Constraints on models}
\label{sec:phenomodels}

The new constraints that we have derived 
will generally place strong bounds on the couplings
of a new light vector. In this section, we illustrate
this by considering a number of models from the literature,
beyond the baryon number vector and $U(1)_R$ models
discussed in earlier sections.

To avoid constraints from neutrino couplings,
some models introduce purely right-handed couplings
to fermions. An example is the $\mu_R$ model of~\cite{Batell:2011qq},
which introduces a light vector with generation-specific
couplings to right-handed muons, motivated by anomalies in low-energy muon physics.
This leads to $X Z \gamma$ and $X \gamma \gamma$ anomalous couplings,
and LHC $Z \rightarrow \gamma X$ searches should be able 
to place stronger
constraints on the coupling than the processes considered in~\cite{Batell:2011qq,Aaij:2016qsm}. Furthermore, the $\mu_R-\tau_R$ version of the same model 
would avoid $Z\to \gamma X$ constraints, but could be probed via $ e^+e^-\to \tau^+\tau^- X$ processes at $B$-factories, as discussed further in section~\ref{sec:directions}.

For an example featuring right-handed couplings to quarks and leptons,
the model of~\cite{Kahn:2007ru} proposes a light ($10- 100 \MeV$)
vector with axial couplings to first-generation fermions,
which would affect the rare $\pi^0 \rightarrow e^+ e^-$ decay.
One point to note is that, while they only demand that
the electron couplings are mostly right-handed (to suppress
neutrino constraints), the down-type quark couplings should also be right-handed
to avoid significant extra constraints from FCNCs,
as discussed in section~\ref{sec:penguin}.
If we consider right-handed couplings to all fermions, then unless
the couplings are chosen to cancel anomalies in the EFT,
anomalous production constraints will give strong bounds.
Doing this requires an electron coupling $\sim 3$ times larger
than their fiducial parameters, making the model more constrained.
This logical sequence illustrates some of the non-obvious requirements
on light vector models.

It has recently been claimed that measurements of
$^8$Be decays provide evidence
for the existence of a new light vector of mass $\simeq 17 \MeV$,
coupling to electrons and nucleons~\cite{Krasznahorkay:2015iga}.
A number of papers have attempted to construct models
for such a vector~\cite{Feng:2016ysn,Kozaczuk:2016nma,Gu:2016ege,Seto:2016pks,Neves:2016ugb,Neves:2017rcn,DelleRose:2017xil,Cao:2017cze}.
Since neither a dark photon or a $B-L$ vector can account for the data
(the required couplings are excluded by other constraints, such as $\pi^0$ 
Dalitz-type decays and neutrino scattering),
it provides an interesting case study. Here, we point out that a variety
of the models in the literature are ruled out by our new constraints, unless
extra fine-tuning at or above the weak scale is introduced:

\begin{itemize}
	\item The baryon number vector model of~\cite{Feng:2016ysn} requires a coupling $g_X \gtrsim 6\times 10^{-4}$ paired with a large kinetic mixing 
$\epsilon \gtrsim 10^{-3}$. These parameters result in ${\rm Br}(B \rightarrow K^* X) \simeq 2 \times 10^{-4}$ from the anomalous $XWW$ coupling, well above
		the experimental bound of $\Delta {\rm Br}(B \rightarrow K^* e^+ e^-) \lesssim 10^{-6}$
		(see section~\ref{sec:fcncexp}).
	\item The $B-L$ model of~\cite{Feng:2016ysn} has the SM leptons
		and neutrinos mix with new fields, giving the light mass eigenstates
		(the physical leptons and neutrinos) altered couplings to a $B-L$ boson.
		This allows the authors of~\cite{Feng:2016ysn} to avoid the stringent bounds coming from
		neutrino-electron scattering, by suppressing the coupling of the
		new vector to neutrinos. However, this weak-isospin breaking in the
		EFT means that $\pi^\pm \rightarrow e^\pm \nu X$ decays,
		as discussed in section~\ref{sec:ccrad}, are energy-enhanced.
		In the model of~\cite{Feng:2016ysn}, in which $X$ has dimension-4 couplings to electrons, but not to pions or neutrinos,
the induced $\pi^+ \rightarrow e^+\nu_e X$ branching ratio is
		\begin{equation}\label{piondecay}
{\rm Br}_{\pi^+ \rightarrow e^+\nu_e X} \simeq 1.5 \times 10^{-9}
\left(\frac{g_{eX}}{10^{-4}}\right)^2 \left(\frac{17 \MeV}{m_X}\right)^2\,,
\end{equation}
		while they require $g_{eX} > 3 \times 10^{-4}$
		to account for the claimed anomaly.
		As per section~\ref{sec:ccrad}, the experimental constraint
		on this branching ratio, for $X \rightarrow e^+ e^-$,
		is $\lesssim 10^{-9}$. The combination of bounds such as equation (\ref{piondecay}) and neutrino scattering constraints is particularly 
		powerful in constraining $B-L$-based models,
		since they are both effects in the low-energy theory,
		which cannot be fine-tuned away by UV physics
		(in particular, neutrino-electron scattering is measured over
		a range of momentum transfers, so cannot be cancelled by a contact operator).

		In addition, since the new fermions which mix with the SM 
		states are heavy, integrating them out gives an anomalous EFT,
		with an effective $\propto \frac{\varphi}{f_X} (g^2 W^a \tilde{W}^a - g'^2 B \tilde{B})$
		ALP-like anomalous coupling. Though they envisage the new fermions
		having mass $\sim 100 \GeV$, which is not heavy enough to be entirely
		unimportant in $Z$ decays or FCNC meson decays, an approximate
		calculation shows that their fiducial parameters
		should be fairly comfortably ruled out by 
		$B \rightarrow K^* X$ decays.

	\item Ref.\ \cite{Kozaczuk:2016nma} proposes a light vector with a
		combination of (generation-universal) axial and vector couplings to quarks.
		At the level of nuclear amplitudes, the axial-current induced transitions
		are enhanced compared to vector-current induced transitions by the ratio of the proton mass to the energy of the transition.
		Consequently, parametrically smaller axial-vector couplings can result in the
		 same strength nuclear transitions, and the authors of \cite{Kozaczuk:2016nma} find a small window of allowed couplings.
		 The authors find that couplings of $ {\cal O} ( 10 ^{ - 5} ) $ can explain the excess. However, flavour-universal axial-vector couplings of this strength are in tension with 
		our meson decay bounds, as per the results
		of previous sections.
		In particular, the coupling to right-handed down-type quarks is problematic for $\Upsilon$ decay bounds (which are not UV-dependent), 
	while the couplings to left-handed quarks, and right-handed up-type quarks, are severely constrained by FCNC
	$B$ decays, in the absence of extra fine-tuning.
	\end{itemize}

While constraints of these kinds will generally apply to light vector
models which could explain the claimed $^8$Be anomaly, 
it is possible that models with additional fine-tuning of the
troublesome amplitudes (e.g.\ fine-tuning of FCNC amplitudes via some carefully constructed initial
quark non-universality of $X$ couplings) could be found.
From this perspective, 
it is important to test the credibility of the experimental
claim~\cite{Krasznahorkay:2015iga} independently of new physics
models, and, if correct, scrutinize possible nuclear physics
effects that could create anomalies in the angular distribution of the
electron-positron pair \cite{Zhang:2017zap}.


\section{Future experimental directions} 
\label{sec:directions}

As we have demonstrated, the dominant
production processes for new light vectors
coupled to non-conserved SM currents can be very different
to those for e.g.\ a dark photon.
These processes allowed us to place more stringent bounds
on the couplings of such vectors from existing experimental data.
They also suggest new classes of observations, which would
allow future searches to significantly improve
their discovery and exclusion potential.
Here, we summarise some of the new searches
and signatures which could be utilised by experimental collaborations:

\begin{itemize}

	\item {\emph {$B$ meson decays at $B$ factories and LHCb:}}
		FCNC $B \rightarrow K^{(*)} X$ meson decays are one of the most
		powerful probes of many light vector models.
		For a hadronically coupled $X$, if $m_X \gtrsim m_\pi$ then
		it can decay to $\pi^0 \gamma$ or $\pi^+ \pi^- \pi^0$,
		as well as higher-multiplicity pion or kaon final states,
		or nucleon-antinucleon pairs, for larger masses. 
		These would result in peaks
		in the invariant mass distribution for the corresponding
		combination of the final state hadrons at $m_X$.
		Given the capability of $B$-factory experiments
		to resolve invariant masses with charged hadrons, $\gamma$, and $\pi^0$,
		such ``bump hunt'' analyses of $B \rightarrow K^{(*)} + {\rm hadronic}$
		(and electromagnetic) final states can be done with existing
		data from BaBar and Belle.
		Such searches could also be added to the Belle-II program.

		$B \rightarrow K^{(*)} X$ decays in which $X$ decays leptonically
		can also be probed at $B$ factories.
		Currently, the only published analyses
		in the $m_{ee} \lesssim m_{\pi^0}$ regime are for the $B \rightarrow K^* e^+ e^-$ decay,
	 since there is interest in the low-$q^2$ enhancement
		of the SM rate. However, for our purposes, the fact that
		the SM rate for $B \rightarrow K e^+ e^-$ is not enhanced at low $q^2$
		is actually helpful for finding new physics contributions.
		In general, a fine-binned search in the $l^+ l^-$
		mass spectrum in such decays would significantly tighten
		bounds on light, leptonically decaying $X$.
		In addition to $B$-factory searches, LHCb
		also has the ability to set constraints of this kind.

		In addition to FCNC decays, leptonic decays
		of the form $B^\pm \rightarrow l^\pm \nu + l^+ l^-$
		occur via
		$B^\pm \rightarrow l^\pm \nu X$, and are enhanced if $X$
		has weak-isospin violating couplings to leptons.
		Unlike the corresponding pion and kaon decays (section~\ref{sec:ccrad}),
		the $B^\pm \rightarrow l^\pm \nu$ decays already have a very
		small SM branching ratio, and it is not clear whether
		a new physics contribution from realistically small $X$ couplings
		would be visible.

	\item {\emph {$\tau\bar\tau X$ production at $B$ and tau/charm factories:}}
	A unique capability of $B$-factories is the possibility
to study $\tau^+\tau^-$ pairs in a well-controlled environment. 
If $X$ has an axial coupling to the $\tau$, then
the $e^+ e^- \rightarrow \tau^+ \tau^- X$ process gives $(m_\tau/m_X)^2$ enhanced
$X_L$ emission, in analogy to the
$e^+ e^- \rightarrow \tau^+ \tau^- S$ signatures discussed in~\cite{McKeen:2011aa,Batell:2016ove}.
Sensitivity estimates show that BaBar and Belle searches
should be able to probe
$|c^A_t| g_X/m_X \lesssim (150 \GeV)^{-1}$ for $200 \MeV \lesssim m_X \lesssim 4 \GeV$~\cite{Batell:2016ove}, for leptonic $X$ decays.

	\item {\emph {Kaon decays at NA62:}} The NA62 kaon 
		experiment~\cite{Anelli:2005ju} at CERN 
		will give improved bounds on charged kaon decays.
		These include $K^+ \rightarrow \pi^+ X$ FCNC decays,
		which will be constrained by
		$K^+ \rightarrow \pi^+ + $
		missing energy and $K^+ \rightarrow \pi^+ l^+ l^-$ searches,
		and $K^+ \rightarrow e^+ \nu X$ decays (section~\ref{sec:ccrad}),
		constrained by $K^+ \rightarrow e^+ \nu l^+ l^-$ searches.

	\item {\emph {Rare $Z$ and $W$ decays at the LHC:}}
		It it well-known that LHC searches for rare EW boson decays can provide
		good sensitivity to exotic states 
		(see e.g.\ Refs.
		\cite{Altmannshofer:2014pba,Izaguirre:2015pga,Elahi:2015vzh}), not least
		due to the extremely large number of $W$s and $Z$s produced
		at the LHC.
		  Besides the $Z\to l^+ l^- l'^{+}l'^{-}$ mode, which already imposes
		important constraints on dark photon and $L_\mu-L_\tau$ models, both
		collaborations should analyze $\gamma X$ final states, with $X$ decaying
		to $l^+l^-$, $\pi^0 \gamma$, $\pi^0\pi^+ \pi^-$ etc.
		Such decays provide strong sensitivity to light vectors
		with anomalous $Z\gamma X$ couplings, such as a baryon number
		coupled vector,
		and should eventually be able to supersede LEP-derived bounds
		(except for the $Z \to \gamma + {\rm invisible}$ channel).
		For $X$ with a weak-isospin-violating difference in couplings
		to leptons and neutrinos, 
		$W \to l \nu X$ decays may also provide constraints.
		If $X$ decays leptonically, then a bump hunt in the invariant
		mass of the $X$ decay products could offer good sensitivity.

		LHC processes other than EW boson decays may also
		be important in some models, and can have longitudinal emission
		enhanced by parametrically higher energies.

	\item {\emph {Displaced decays:}} a number of proposed and upcoming
		experiments will have improved sensitivity to very displaced
		decays of a light vector. The SHiP proton beam dump
		experiment~\cite{Alekhin:2015byh} can significantly improve
		sensitivity to weakly-coupled $X$ produced in $B$-meson decay
		processes. For lower-mass $X$s produced in kaon decays,
		neutrino-related experiments such as the new / planned liquid argon
		near detectors at Fermilab~\cite{Acciarri:2016smi} should achieve
		excellent discrimination of final state products, 
	    potentially giving increased sensitivity. At higher energies, the
		MATHUSLA proposal to build a specialized detector capable of finding
	  	very displaced
		decays at the LHC~\cite{Chou:2016lxi} could improve bounds
		on light, weakly-coupled vectors, though it is not clear whether
		there is viable parameter space in which longitudinal production
		is the dominant process.

\end{itemize}

\section{Discussion}
\label{sec:disc}

The physics point underlying this paper is that, for new vector particles
with dimension-4 couplings to the SM, production of the vector's
longitudinal mode is generally enhanced by ($E$/vector mass)$^2$,
where $E$ is some energy or mass scale associated with the SM process.
The only way to avoid this entirely is to couple to a conserved SM current,
i.e.\ $B-L$ or EM. Practically, coupling to special leptonic currents
such as $L_\mu - L_\tau$ also makes longitudinal production negligible, since
the non-conservation of the current is controlled by the small neutrino
masses. Purely right-handed couplings to first-generation quarks
can also have longitudinal production suppressed by small quark masses.
Energy-enhanced longitudinal mode production
is a well-known effect, but as we have noted, it has been ignored in
many works which consider light BSM vector particles. We have 
discussed a number of cases in which such energy-enhanced production
leads to significantly stronger constraints than derived in existing
literature, and pointed out how these can rule out a number of
phenomenologically-motivated models. As well as setting constraints, 
our work highlights how future measurements and analyses
may allow high-energy experiments to be even more sensitive
probes of weakly-coupled physics.


\begin{acknowledgments}
We thank Masha Baryakhtar, Emilian Dudas, Felix Kahlhoefer, Yotam Soreq, Jesse Thaler, Felix Yu, and Yue Zhao for helpful discussions.
We thank Felix Yu for pointing out a factor-4 error in our original equation~\ref{GaZ}.
Research at Perimeter Institute is supported by the Government 
of Canada through Industry Canada and by the 
Province of Ontario through the Ministry of Economic 
Development \& Innovation. JD is supported in part by the NSF through grant PHY-1316222. 
\end{acknowledgments}


\appendix


\section{Constraints on new SM-chiral fermions}
\label{ap:chiral}

To cancel electroweak anomalies with new SM-chiral fermions getting
their mass from the SM Higgs, we would need at least two new $SU(2)_L$ doublets.
\cite{Bizot:2015zaa} considers the EWP and Higgs decay constraints on
additional chiral fermions (see also~\cite{deGouvea:2012hc}), finding
that more then two new doublets are ruled out, but that two new doublets (along
with four new hypercharged singlets as their partners) are marginally
allowed, for suitable choices of the hypercharge.

For these allowed hypercharge assignments, the new states are all EM
charged. The only way for these charges to be integers is for the states
to have charges 1 and 2. This leads to a $\sim 10\%$ increase in the
$h \rightarrow \gamma \gamma$ width (by flipping the sign of the amplitude),
and a $\sim 130\%$ increase in the $h \rightarrow Z \gamma$ width. 
With $300 \fbm$ of LHC data, the projections of the fractional measurement
accuracy
for these decay channels are $\sim 0.14$ and $\sim 0.45$ 
respectively~\cite{atlas2014projections}, so such deviations
should be detected or ruled out (see also~\cite{No:2016ezr}
for proposed methods of improving the $Z\gamma$ measurement). 
This
is true more generally --- deviations will show up in either $h
\rightarrow \gamma \gamma$ or $h \rightarrow Z \gamma$ for any choice of
the hypercharge.

If we also consider direct production signals, it is likely that any
such model would already have been seen at the LHC.
If the new fermions
have non-integer EM charges (in which case
at least one must have charge $\gtrsim 0.8$), then they are stable.
As well as the problems arising from a possible cosmological relic
population of stable non-integer-charged states, there
are LHC constraints on such particles,
which gives a lower mass limit of $\sim 700 \GeV$~\cite{CMS:2016ybj}.
For the case where the charges are 1 and 2, the $Q=2$ fermion can
decay to a $W$ boson and the $Q=1$ fermion, while the latter
can mix with the SM leptons, and decay to $Zl$ or $hl$.
\cite{Ma:2014zda} considers the signatures of these kinds of new fermions, 
finding that $\sim 100 \fbm$ of $14 \TeV$ data should
place a mass limit of $\gtrsim 500 \GeV$.
To make the new chiral fermions heavier than these
lower bounds, we would need to give them very large Yukawa couplings,
which introduces strong coupling problems.

In summary, the next LHC run should, if it finds no deviations
from the SM, fairly robustly exclude next chiral fermions
obtaining their mass from the SM Higgs,
and even now, constructing models with such fermions would be
a very delicate task.


\section{Anomalous amplitudes}
\label{ap:anom}

If $\chi_i$ are massless chiral (left-handed) fermions which couple to vector
particles $V_j$, with $\LL \supset g_{ij} V_j^\mu \bar\chi_i \gamma_\mu \chi_i$,
then the longitudinal amplitude for the $V_1 V_2 V_3$ triangle 
diagram is
\begin{gather}
	- (p + q)_\mu {\cal M} ^{\mu \nu\rho}
	= 
	\frac{1}{12 \pi^2} \epsilon^{\nu\rho\lambda\sigma} p_\lambda q_\sigma
	\sum_i g_{i1} g_{i2} g_{i3} \,,\notag
 \\[0.5ex] 
	\hspace{0.5cm}	{\cal M} ^{\mu\nu\rho} \equiv 
	\sum_i\hspace{-0.2cm}\adjustbox{valign=m}{
 \begin{tikzpicture}[line width=0.75] 
\coordinate (C1) at (.75,0);
\coordinate (C2) at (0.75+0.75,{0.75*0.7});
\coordinate (C3) at (0.75+0.75,-{0.75*0.7});
\coordinate (C4) at (0.75+0.75+0.75,{0.75*0.7});
\coordinate (C5) at (0.75+0.75+0.75,-{0.75*0.7});
    \draw[v] (0,0) node[left]{$ V_1 ^\mu  $} -- (C1);
    \draw[f] (C1) -- (C2);
    \draw[f] (C2) -- (C3) node[midway,right] {$ i $};
    \draw[fb] (C1) -- (C3);
    \draw[v] (C2) -- (C4) node[right] {$ V_2 ^\nu  $} node[above,midway] {$  p \rightarrow  $};
    \draw[v] (C3) -- (C5)node[right] {$ V_3 ^\rho  $}node[below,midway] {$  q \rightarrow  $}; 
 \end{tikzpicture}} \label{eq:chiral1} \,,
\end{gather}
for any regularisation method that respects the symmetry
between the three legs. This is termed 
the `consistent anomaly'~\cite{Bilal:2008qx}, with $p_\nu \MM^{\mu\nu\rho}$
and $q_\rho \MM^{\mu\nu\rho}$ being given by the obvious symmetrical
expressions. For vectors whose couplings
are non-diagonal, such as non-abelian gauge bosons,
the triangle amplitude
is given by the obvious modification of equation~\ref{eq:chiral1},
where there must be no overall fermion flavor change around the loop.

In many circumstances, it is helpful to use a regularisation
method which does not treat the legs symmetrically.
For example, the gauge symmetries corresponding to 
some of the vectors may be broken, but others preserved.
If we evaluate the $XBB$ triangle diagram below,
using a regulator which preserves
the $U(1)_B$ gauge symmetry, we obtain the `covariant anomaly'
\begin{gather}
	- (p + q)_\mu \MM^{\mu \nu\rho} (-p-q,p,q) 
	= 
	\frac{1}{4 \pi^2} \epsilon^{\nu\rho\lambda\sigma} p_\lambda q_\sigma
	\sum_i g_{iX} g_{iB}^2\,,
	\label{eq:covar1} \\
	p_\nu \MM^{\mu\nu\rho} = 0  \,, \quad 
	q_\rho \MM^{\mu\nu\rho} = 0 \,,
	\\[0.5ex]
	\hspace{0.5cm}	{\cal M} ^{\mu\nu\rho} \equiv 
	\sum_i\hspace{-0.2cm}\adjustbox{valign=m}{
 \begin{tikzpicture}[line width=0.75] 
\coordinate (C1) at (.75,0);
\coordinate (C2) at (0.75+0.75,{0.75*0.7});
\coordinate (C3) at (0.75+0.75,-{0.75*0.7});
\coordinate (C4) at (0.75+0.75+0.75,{0.75*0.7});
\coordinate (C5) at (0.75+0.75+0.75,-{0.75*0.7});
    \draw[v] (0,0) node[left]{$ X _\mu  $} -- (C1);
    \draw[f] (C1) -- (C2);
    \draw[f] (C2) -- (C3) node[midway,right] {$ i $};
    \draw[fb] (C1) -- (C3);
    \draw[v] (C2) -- (C4) node[right] {$ B _\nu  $} node[above,midway] {$  p \rightarrow  $};
    \draw[v] (C3) -- (C5)node[right] {$ B _\rho  $}node[below,midway] {$  q \rightarrow  $}; 
  \end{tikzpicture}} \notag \,.
\end{gather}
In a regularisation scheme which is symmetric between $X$ and $B$,
we can obtain the `covariant' result by including an explicit 
Wess-Zumino term
\begin{equation}
	\LL \supset \frac{1}{6 \pi^2} \epsilon^{\mu\nu\rho\sigma} X_\mu B_\nu \partial_\rho B_\sigma
	\sum_i g_{iX} g_{iB}^2\,.
\end{equation}
Within a UV completion in which both the $U(1)_X$ and $U(1)_B$
symmetries are only spontaneously broken,
this WZ term will arise from integrating out heavy anomaly-cancelling
fermions, as illustrated in section~\ref{sec:xlamps}.

We can find explicit examples of these different types of regulators
by considering different forms of Pauli-Villars (PV) regularisation.
For the covariant anomaly, if the fermions in the loop have
$B$-preserving but $X$-breaking Dirac masses, then
we can regulate by introducing one heavy Dirac PV fermion for each
of the Dirac fermions in the original theory. Since the couplings
of the PV fermions preserve $B$, we obtain the covariant amplitude
(in the limit where the original fermions have very small masses
--- we will return to the mass dependence of the amplitude below).

To obtain the consistent anomaly, we can introduce a PV Dirac fermion
for every Weyl fermion in our original theory (rather than one per pair
of original fermions), with only the left-handed parts of each PV Dirac
fermion coupling to the vectors. This procedure breaks all of the
gauge symmetries, unless the anomalies cancel when summing over chiral
fermions. We can restore some of these symmetries,
and recover the corresponding covariant anomaly, using appropriate WZ terms.
This scheme has the advantage of applying unchanged in different phases
of a theory, as illustrated in section~\ref{sec:xlamps}.

For vectors with non-diagonal couplings to the chiral fermions,
as occurs with non-abelian gauge bosons, there can be
four-leg (`box') and five-leg (`pentagon') anomalous amplitudes,
as well as the triangle amplitudes discussed above.
For the longitudinal amplitudes
of an abelian vector (with diagonal couplings) against other non-abelian
vectors, the pentagon diagrams are not anomalous~\cite{Bilal:2008qx},
leaving only the triangle and box diagrams.
A new vector coupling to a tree-level conserved current in the SM,
as considered in section~\ref{sec:anom}, gives an example of the latter,
with the $X-SU(2)_L-SU(2)_L$ anomaly giving $XWWW$ couplings,
coming from box diagrams.


\subsection{Fermion mass dependence}

Since chiral anomalies can be derived from topological 
considerations~\cite{Bilal:2008qx}, they are independent of fermion masses. However, we are interested
in longitudinal mode production, which corresponds to the divergence
of the current that our vector couples to. When fermions have mass terms which
break $U(1)_X$, the variation of the action under $U(1)_X$ 
receives contributions from these mass terms, so the divergence is the sum
of these mass-dependent pieces plus the mass-independent anomaly. 

For example, suppose that $X$ couples to Weyl fermions $\psi_L$ and
$\psi_R$, which are the left-handed and right-handed components of a
Dirac fermion with mass term $\LL \supset - m \bar{\psi} \psi $ (in four-component
notation). Then, we have the operator equations
\begin{align}
& 	\partial_\mu (\bar\psi_L \gamma^\mu \psi_L) + i m \bar\psi \gamma _5  \psi = {\rm anomaly} \,,\\ 
& 	\partial_\mu (\bar\psi_R \gamma^\mu \psi_R) + i m \bar\psi \gamma _5  \psi = {\rm anomaly}\,,
      \end{align}
      where `anomaly' denotes the product of field strength tensors arising from the anomaly. If
      \begin{equation}
        {\cal L}  \supset X_\mu (g_L \bar\psi_L \gamma^\mu \psi_L + g_R \bar\psi_R \gamma^\mu \psi_R)\,,
      \end{equation}
      then the matrix element between the vacuum and two gauge fields, $ V _1 $ and $ V _2 $, is,
\begin{align}
 &  -(p + q)_\mu {\cal M} ^{\mu\rho\sigma} \notag \\
	&\hspace{0.2cm}=  \braket{ V _1 ^\rho V _2 ^\sigma | g_L\partial_\mu \bar\psi_L \gamma^\mu \psi_L + g _R \partial_\mu \bar\psi_R \gamma^\mu \psi_R  | 0 }	\,.
\end{align}
This can be rewritten using the operator equation as,
\begin{equation}
= (g_R - g_L) i m \langle  V _1 ^\rho V _2 ^\sigma | \bar\psi \gamma _5  \psi | 0 \rangle + {\rm anomaly}\,.
  \end{equation} 
The mass-dependent contribution is therefore set by the axial coupling
of $X$ to $\psi$, and the vacuum-to-two-vector matrix element of
the $\bar\psi \gamma_5 \psi$ operator.
This can also be seen by direct calculation of vector triangle diagrams with
massive Dirac fermions, e.g.~\cite{Hill:2006ej}.
In cases where we are considering new fermions with large, non-EW-breaking
masses, such as section~\ref{sec:xlamps}, 
we are interested in e.g.\ $X_L B B$ amplitudes, where $B$
couples vectorially to the massive Dirac fermions in the loop.
If $X$ couples to a given Dirac fermion $\psi$ as $\LL \supset X_\mu
\bar\psi (g_V + \gamma_5 g_A) \gamma^\mu \psi$, then for a `consistent'
regulator we have,~\cite{AdlerAnomalies,Schwartz:2013pla}
	\begin{equation}
	- (p + q)_\mu \MM^{\mu \nu \rho}= 
	 \frac{g_X g^{\prime 2}}{2 \pi^2}
	\epsilon^{\nu\rho\lambda\sigma} p_\lambda q_\sigma
	\left(\frac{1}{3} + 2 m _f ^2 I_{00}(m _f , p,q) \right)\,,
	\end{equation}
where the $1/3$ term is from the anomaly and the mass term gives
\begin{widetext}
  \renewcommand{\arraystretch}{1.3}
  \begin{equation}
 2 m ^2 _f  	I_{00}(m _f , p,q)
	\equiv \int_0^1 dx \int_0^{1-x} dy\, \frac{2 m_f^2}{y(1-y) p^2 + x(1-x) q^2 
	+ 2 x y \, p \cdot q - m^2 _f }
	\sim 
\left\{ 		\begin{array}{cl}
	 -  1 & \hspace{0.15cm}, \, \, m _f ^2 \gg p^2, q^2, p \cdot q \\
			0 & \hspace{0.15cm}, \, \, m^2 _f  \ll p^2, q^2, p \cdot q 
		\end{array} \right. \,.
	\end{equation}
\end{widetext}
For large masses, the mass and the anomaly contributions combine to give the net divergence of the non-conserved current. In particular, in the limit of large $m$, 
\begin{equation}
	- (p + q)_\mu {\cal M} ^{\mu \nu \rho}\simeq 
	- g_X g^{\prime 2} \frac{1}{3 \pi^2}
	\epsilon^{\nu\rho\lambda\sigma} p_\lambda q_\sigma\,.
\end{equation}
This is minus the contribution from the $B$-covariant WZ term.
So, for a $B$-covariant regulator, the longitudinal $X$ coupling
from such heavy fermions is suppressed by powers of $1/m^2$,
as required by decoupling.
A similar story applies for anomalous box diagrams.

As is well-known~\cite{Adler:2004qt}, the anomaly
is given exactly by the 1-loop calculation (and the same
applies to the matrix element of the $\bar\psi \gamma_5 \psi$ operator).


\section{Tree-level FCNCs}
\label{ap:treefcnc}

The non-diagonal quark and neutrino Yukawa matrices in the SM mean
that, even if there are no tree-level flavor-changing $X$ vertices
before EWSB, they may be induced by the SM fermions getting masses.
In the SM interaction basis for the quarks, we can write the Lagrangian
after EWSB as
\begin{align}
	\LL &\supset
	\bar{u}_R^i \left(i \slashed{\partial} + g' \frac{2}{3} \slashed{B} + g_X Q^u_{ij} \slashed{X} \right) u_R^j \\
	&+ 
	\bar{d}_R^i \left(i \slashed{\partial} - g' \frac{1}{3} \slashed{B} + g_X Q^d_{ij} \slashed{X} \right) d_R^j \notag \\
	&+ 
	\bar{Q}_L^i \left(i \slashed{\partial} + \dots + g_X \slashed{X}
	\begin{pmatrix} Q^U_{ij} & 0 \\ 0 & Q^D_{ij} \end{pmatrix}\right) Q_L^j \notag \\
		&+ \left( \bar{d}_L^i (U_d M_d K_d^\dagger)_{ij} d_R^j
		+ \bar{u}_L^i (U_u M_u K_u^\dagger)_{ij} u_R^j + {\rm h.c.}
		\right) \,,\notag
\end{align}
where $M_u$ and $M_d$ are the diagonal quark mass matrices,
$U_q$ and $K_q$ are unitary matrices, and $V \equiv U_u^\dagger U_d$
is the CKM matrix. Without loss of generality, $Q^u$ and $Q^d$ are diagonal (after a rotation
of $u_R^i$ and $d_R^i$). While $K_u$ and $K_d$ can be rotated away in the SM,
here they have physical consequences if $Q^u$ and $Q^d$ are not the identity.
However, since SM measurements are compatible with $K_u = K_d = 1$,
it is possible for the $X$ couplings to right-handed fermions to be diagonal
in the mass basis, without being generation-universal.

However, the CKM matrix elements can be measured in the SM, and $V
\neq 1$. Thus, if we demand diagonal couplings in the mass basis, then
$Q^U$ and $Q^D$ can only be proportional to each other if they are both
proportional to the identity, or if one of them is zero.
Hence, if there are generation-non-universal couplings to both
up-type and down-type quarks, then there must be weak-isospin
breaking to avoid tree-level FCNCs.

If tree-level FCNC vertices are present, the constraints on their
couplings are very stringent, since
amplitudes lack the loop and coupling suppressions
of those discussed in section~\ref{sec:penguin}. To give
an example, if we want a vector to couple dominantly to first-generation
quarks, and to have non-zero $u_L$ and $d_L$ couplings, then
at least one of the $sdX$ and $cuX$ vertices must have
coupling $\sim g_X \theta_c$, where $\theta_c \sim 0.2$ is the Cabbibo angle.
The $K \rightarrow \pi X$ decay rate from the $sdX$ vertex is roughly 
\begin{equation}
	{\rm Br}(K \rightarrow \pi X) \sim 10^{-5} \left(\frac{|g_{sdX}|}{10^{-10}}\right)^2 \left(\frac{100 \MeV}{m_X}\right)^2
\end{equation}
so experimental constraints would require that $g_X$ is extremely small.

If neutrinos are Majorana, then it is always possible that the
lepton Yukawas are such that $X$ has flavor-diagonal couplings in
the lepton mass basis. If they are Dirac, then a similar analysis
to the above discussion for quarks applies.


\bibliography{lightVectors}
\end{document}